\documentclass[preprint,superscriptaddress,aps,pre]{revtex4}
\usepackage{graphicx,amsmath} 
\date{\today}

\begin{document}

\title{Non-Markovian theory of vibrational energy relaxation
and its applications to biomolecular systems}


\author{Hiroshi FUJISAKI}\email{fujisaki@nms.ac.jp}
\affiliation{
Department of Physics,
Nippon Medical School,
2-297-2 Kosugi-cho, Nakahara, Kawasaki 
211-0063, Japan
}
\affiliation{
Molecular Scale Team, Integrated Simulation of Living Matter Group, 
Computational Science Research Program, 
RIKEN, 
2-1 Hirosawa, Wako, Saitama 351-0198, Japan
}
\author{Yong ZHANG}\email{zhangy1977@gmail.com}
\affiliation{
Center for Biophysical Modeling and Simulation and Department of Chemistry, 
University of Utah, 315 S. 1400 E. Rm. 2020, Salt Lake City, Utah 84112-0850, 
USA}
\author{John E. STRAUB}\email{straub@bu.edu}
\affiliation{
Department of Chemistry, Boston University,
Boston, Massachusetts, 02215, USA
}






\maketitle

\section{Introduction}

Energy transfer (relaxation) phenomena are ubiquitous in nature. 
At a macroscopic level, the phenomenological  
theory of heat (Fourier law) successfully describes heat transfer and energy flow. 
However, its microscopic origin is still under debate. 
This is because the phenomena can contain many-body, multi-scale, 
nonequilibrium, and even quantum mechanical aspects, which present 
significant challenges to theories addressing energy transfer 
phenomena in physics, chemistry and biology \cite{Johnbook}. 
For example, heat generation and transfer in nano-devices is a critical problem 
in the design of nanotechnology.
In molecular physics, 
it is well known that vibrational energy relaxation (VER) 
is an essential aspect of any quantitative description of chemical reactions \cite{Nitzan}. 
In the celebrated RRKM theory of an absolute reaction rate for isolated 
molecules, it is assumed that the intramolecular 
vibrational energy relaxation (IVR) is much faster than the reaction itself. 
Under certain statistical assumptions, 
the reaction rate can be derived \cite{Steinfeldbook}. 
For chemical reactions in solutions, the transition state theory and its extension 
such as Kramer's theory and the Grote-Hynes theory have been developed 
\cite{Billingbook,BBS88} and applied to a variety of chemical systems 
including biomolecular systems \cite{PTMHR08}.
However, one cannot always assume that separation of timescales.
It has been shown that a conformational transition 
(or reaction) rate can be modulated by the IVR rate \cite{LW97}. 
As this brief survey demonstrates, a detailed understanding of IVR or VER 
is essential to study the chemical reaction and conformation change of molecules. 


A relatively well understood class of VER is a single vibrational mode embedded 
in (vibrational) bath modes. 
If the coupling between the system and bath modes is weak (or assumed to be weak), 
a Fermi's-golden-rule style formula derived using 2nd 
order perturbation theory \cite{Oxtoby79,RMH04,LS02} may 
be used to estimate the VER rate. 
However, the application of such theories to real molecular 
systems poses several (technical) challenges, 
including how to choose force fields, how to separate 
quantum and classical degrees of freedom, 
or how to treat 
the separation of timescales
between system and bath modes.
Multiple solutions have been proposed to 
meet those challenges leading to a 
variety of theoretical approaches to
the treatment of VER 
\cite{Okazaki,Leitner,Leitner09,Pouthier08,Knoester07,Knoester08}. 
These works using Fermi's golden rule 
are based on quantum mechanics and 
suitable for the description of high frequency
modes (more than thermal energy $\simeq$ 200 cm$^{-1}$), 
on which nonlinear spectroscopy
has recently focused \cite{Hochstrasser,MK97,Dlott03,Hamm08}.

In this chapter, we summarize our recent work on VER of 
high frequency modes in biomolecular systems.
In our previous work, we have concentrated on the VER rate and 
mechanisms for proteins \cite{Fujisaki05}. 
Here we shall focus on the time course of the VER dynamics. 
We extend our previous Markovian theory of VER to a non-Markovian 
theory applicable to a broader range of chemical systems \cite{FZS06,FS07}. 
Recent time-resolved 
spectroscopy can detect the time course of VER dynamics 
(with femto-second resolution), 
which may not be accurately described by a single time scale. 
We derive new formulas for VER dynamics, and apply them to several 
interesting cases, where comparison to experimental data is available. 

This chapter is organized as follows:
In Sec.~\ref{sec:normalmode}, 
we briefly summarize the normal mode concepts in protein dynamics 
simulations, on which we build our non-Markovian VER theory.
In Sec.~\ref{sec:derivation}, 
we derive VER formulas under several assumptions, 
and discuss the limitations of our formulas.
In Sec.~\ref{sec:application}, 
we apply the VER formulas to several situations: 
the amide I modes in solvated $N$-methylacetamide 
and cytochrome c, and two in-plane modes 
($\nu_4$ and $\nu_7$ modes) in a porphyrin ligated to imidazol. 
We employ a number of approximations in describing the potential
energy surface on which the dynamics takes place, including 
the empirical CHARMM \cite{CHARMM} force field 
and density functional calculations \cite{Gaussian}
for the small parts of the system ($N$-methylacetamide and porphyrin).
We compare our theoretical results with experiment when available, 
and find good agreement. We can deduce the VER mechanism based on 
our theory for each case. In Sec.~\ref{sec:summary}, 
we summarize and discuss the further aspects of VER 
in biomolecules and in nanotechnology (molecular devices).

\section{Normal mode concepts applied to protein dynamics}
\label{sec:normalmode}

Normal mode provides a powerful tool in exploring molecular 
vibrational dynamics \cite{Wilsonbook}
and may be applied to biomolecules as well \cite{CuiBaharbook}.
The first normal mode calculations for a protein were performed 
for BPTI protein \cite{GNN83}. 
Most biomolecular simulation
softwares support the calculation of normal modes
\cite{CHARMM,Gromacs,Amber}.
However, the calculation of 
a mass-weighted hessian $K_{ij}$, which requires the second 
derivatives of the potential energy surface, with elements 
defined as  
\begin{equation}
K_{ij}=\frac{1}{\sqrt{m_i m_j}} \frac{\partial^2 V}{\partial x_i \partial x_j}
\end{equation}
can be computational demanding. 
Here $m_i$ is the mass, $x_i$ is the coordinate, and 
$V$ is the potential energy of the system.
Efficient methods have been
devised including torsional angle normal mode \cite{Wako}, 
block normal mode \cite{TGMS00}, 
and the iterative mixed-basis diagonalization (DIMB) methods \cite{MP93} etc.
An alternative direction for the efficient calculations of a hessian 
is to use 
coarse-grained models such as elastic \cite{Tirion} 
or Gaussian network \cite{HBE97} models.
From normal mode analysis (or instantaneous 
normal mode analysis \cite{Keyes}), the frequencies, density of states, 
and normal mode vectors can be calculated. 
In particular, the latter quantity is 
important because it is known that the lowest eigenvectors
may describe the functionally important motions such as large-scale 
conformational change,  
a subject that is the focus of another chapter in this 
volume \cite{FMK09}.

\begin{figure}[b]
\begin{minipage}{.42\linewidth}
\includegraphics[scale=0.5]{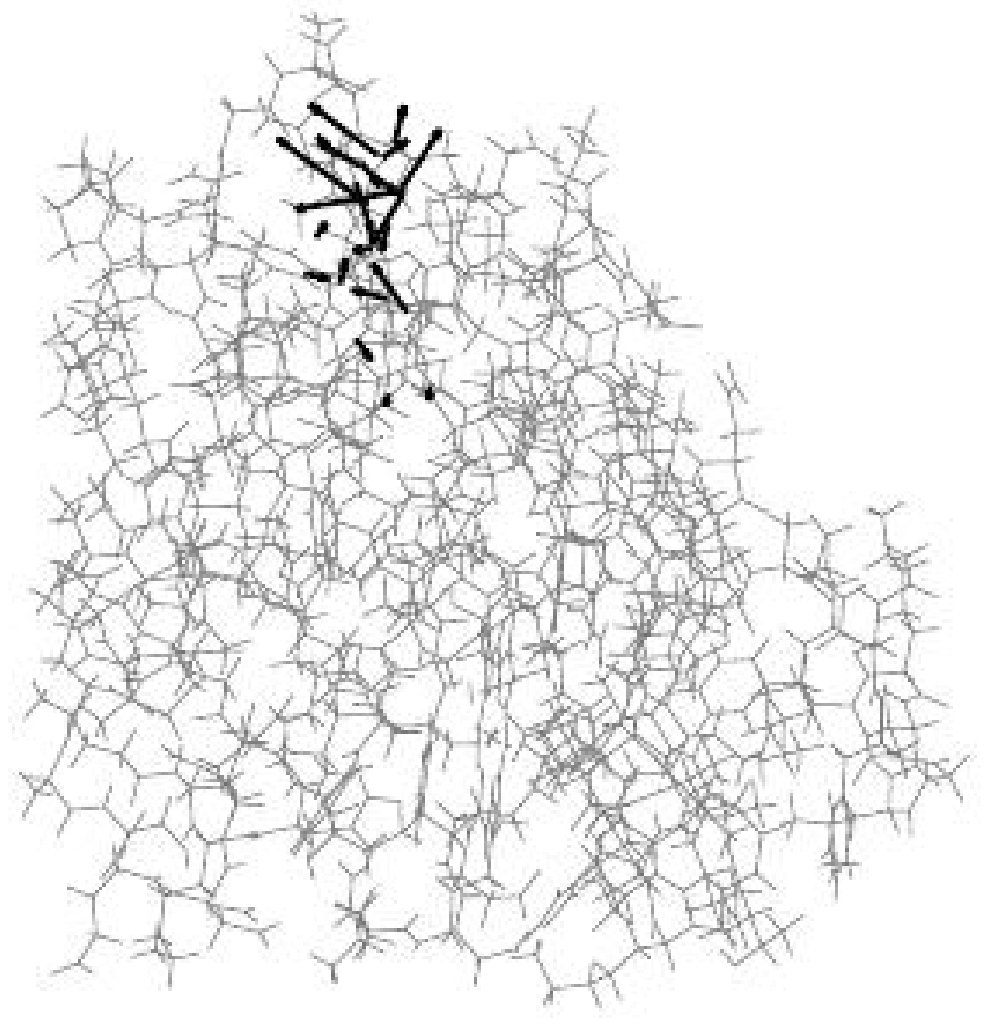}
\end{minipage}
\begin{minipage}{.42\linewidth}
\includegraphics[scale=0.5]{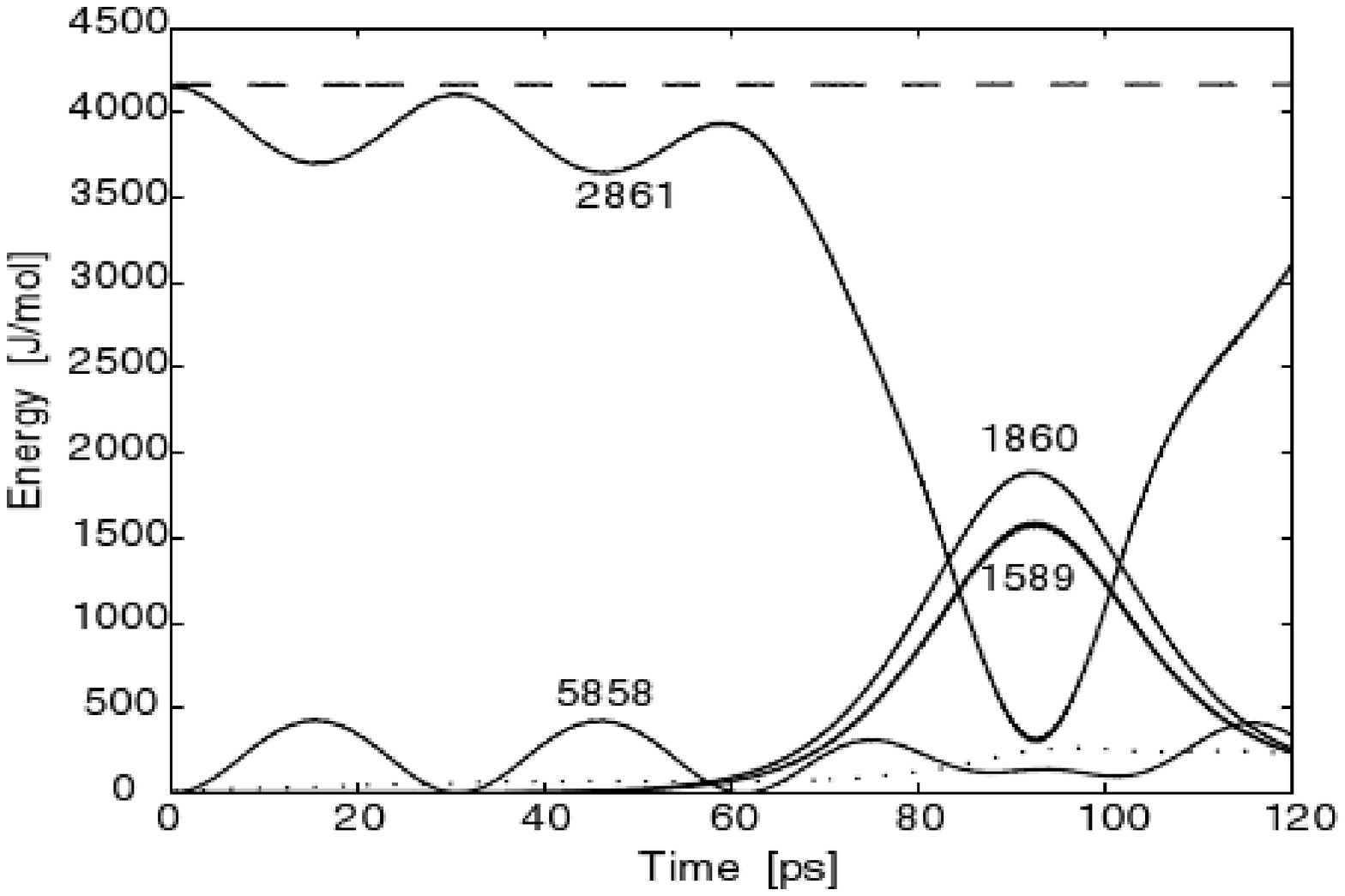}
\end{minipage}
\hspace{1cm}
\caption{\baselineskip5mm
Left: The excited eigenvector depicted by arrows in myoglobin.
Right: Classical simulation of mode specific energy transfer in myoglobin 
at zero temperature. 
[Reprinted with permission from 
K. Moritsugu, O. Miyashita and A. Kidera,
Phys.~Rev.~Lett.~{\bf 85}, 3970 (2000).
Copyright \copyright 2009 by the American Physical Society.]
}
\label{fig:myoglobin}
\end{figure}

There is no doubt as to the usefulness of normal mode concepts.
However, for molecular systems, it is always 
an approximate model as higher-order nonlinear 
coupling and intrinsic anharmonicity become essential.
To describe energy transfer (or relaxation) phenomena in a protein, 
Moritsugu, Miyashita, and Kidera (MMK) introduced a reduced model using 
normal modes with 3rd and 4th order anharmonicity \cite{MMK00}, 
$C_{klm}^{(3)}$ and $C_{klmn}^{(4)}$, respectively,
\begin{equation}
V(\{ q_k \})
=\sum_k \frac{\omega_k^2}{2} q_k^2
+\frac{1}{3!} \sum_{klm} C_{klm}^{(3)} q_k q_l q_m
+\frac{1}{4!} \sum_{klmn} C_{klmn}^{(4)} q_k q_l q_m q_n
\label{eq:reduced}
\end{equation}
with
\begin{eqnarray}
C_{klm}^{(3)} 
&\equiv& \frac{\partial^3 V}{\partial q_k \partial q_l \partial q_m},
\\
C_{klmn}^{(4)}
&\equiv& \frac{\partial^4 V}
{\partial q_k \partial q_l \partial q_m \partial q_n}
\end{eqnarray}
where $q_k$ denotes the normal mode calculated by the hessian $K_{ij}$ 
and $\omega_k$ is the normal mode frequency.
Classical (and harmonic) Fermi resonance \cite{Kubobook}
is a key ingredient in the MMK theory of 
energy transfer derived from observations of all-atom simulations of 
myoglobin at zero temperature (see Fig.~\ref{fig:myoglobin}).

At finite temperature, non-resonant effects 
become important and clear interpretation of the numerical results 
becomes difficult within the classical approximation.
Nagaoka and coworkers \cite{Nagaoka} identified essential
vibrational modes in vacuum simulations of myoglobin and connected these modes to 
the mechanism of ``heme cooling'' explored experimentally
by Mizutani and Kitagawa \cite{MK97}.
Contemporaneously, nonequilibrium MD simulations of solvated 
myoglobin carried out 
by Sagnella and Straub provided the first detailed and accurate 
simulation of heme cooling dynamics \cite{Straub}. 
That work provided support for the conjecture that
the motion similar to those modes 
identified by Nagaoka 
play an important role in energy flow pathways. 


\begin{figure}[b]
\hfill
\begin{center}
\includegraphics[scale=0.6]{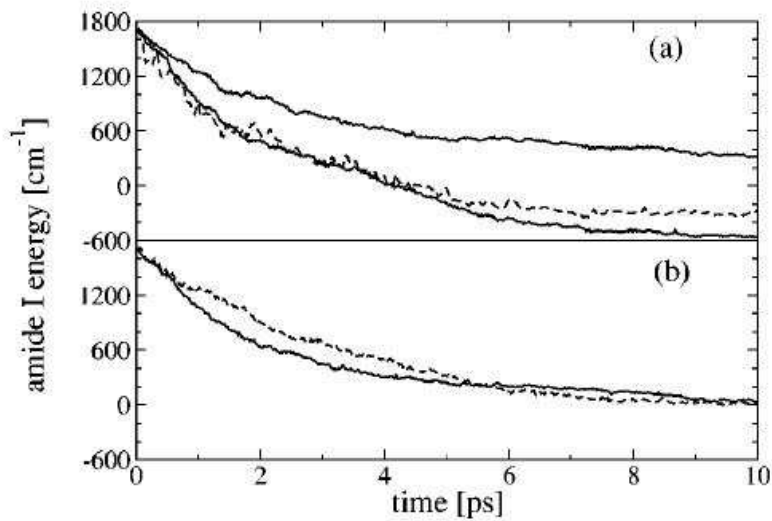}
\end{center}
\caption{\baselineskip5mm
Nonequilibrium MD simulation of energy flow from the excited amide I mode in 
$N$-methylacetamide in heavy water.
See also Fig.~\ref{fig:nma}.
[Reprinted with permission from 
P.H. Nguyen and G. Stock, J.~Chem.~Phys.~{\bf 119}, 11350 (2003).
Copyright \copyright 2009 by the American Institute of Physics.]
}
\label{fig:nguyen}
\end{figure}

Nguyen and Stock explored the vibrational dynamics of the 
small molecule, $N$-methylacetamide (NMA), 
often used as a model of the peptide backbone \cite{Stock}. 
Using nonequilibrium MD simulations of NMA in heavy water, 
VER was observed to occur on a pico-second time scale 
for the amide I vibrational mode (see Fig.~\ref{fig:nguyen}).
They used the instantaneous normal mode concept \cite{Keyes} 
to interpret their result
and noted the essential role of anharmonic coupling.
Leitner also used the normal mode concept to describe 
energy diffusion in a protein, and found an interesting
link between the anomalous heat diffusion and the geometrical
properties of a protein \cite{YL03}.

In terms of vibrational spectroscopy, Gerber {\it et al}. calculated the 
anharmonic frequencies in BPTI, within the VSCF level of theory \cite{Gerber},
using the reduced model (\ref{eq:reduced}). 
Yagi, Hirata, and Hirao refined 
this type of anharmoic frequency calculation for large molecular 
systems with more efficient methods \cite{Yagi}, 
appropriate for applications to biomolecules such as DNA base pair \cite{Yagi2}.
Based on the reduced model (\ref{eq:reduced}) 
with higher-order nonlinear coupling,
Leitner also studied quantum mechanical aspects of VER in proteins,
by employing the Maradudin-Fein theory based on 
Fermi's ``Golden Rule'' \cite{Leitner}. 
Using the same model, Fujisaki, Zhang, and Straub 
focused on more detailed aspects of VER in biomolecular 
systems, and calculated the VER rate, mechanisms or pathways, 
using their non-Markovian perturbative formulas (described below).

As this brief survey demonstrates,
the normal mode concept is a powerful tool that provides 
significant insight into mode specific vibrational 
dynamics and energy transfer in proteins.

\section{Derivation of non-Markovian VER formulas}
\label{sec:derivation}

We have derived a VER formula for the simplest situation, 
a one-dimensional relaxing oscillator coupled to a ``static'' bath \cite{FZS06}.
Here we extend this treatment to two more general directions: 
(a) multi-dimensional relaxing modes coupled to a ``static'' bath
and 
(b) a one-dimensional relaxing mode coupled to a ``fluctuating'' bath \cite{FS08}.

\subsection{Multi-dimensional relaxing mode coupled to a static bath}

We take the following time-independent Hamiltonian
\begin{eqnarray}
{\cal H}
&=& {\cal H}^0_S+{\cal H}_B+{\cal V}^0
\\
&=&
{\cal H}^0_S+\langle {\cal V} \rangle_B
+{\cal H}_B+{\cal V}^0-\langle {\cal V} \rangle_B
\\
&=&
{\cal H}_S
+{\cal H}_B+ {\cal V}
\end{eqnarray}
where
\begin{eqnarray}
{\cal H}_S 
&\equiv& {\cal H}^0_S+\langle {\cal V} \rangle_B,
\\
{\cal V}
&\equiv&
{\cal V}^0-\langle {\cal V} \rangle_B.
\end{eqnarray}
In previous work \cite{FZS06}, we have considered only 
a single one-dimensional oscillator as the system. Here we extend 
that treatment to the case of an $N_S$-dimensional 
oscillator system. That is,
\begin{eqnarray}
{\cal H}_S 
&=& \sum_{i=1}^{N_S}
\left( 
\frac{p_i^2}{2}+\frac{\omega_i^2}{2} q_i^2
\right) +V(\{ q_i \}),
\\
{\cal H}_B 
&=& \sum_{\alpha=1}^{N_B}
\left( 
\frac{p_{\alpha}^2}{2}+\frac{\omega_{\alpha}^2}{2} q_{\alpha}^2
\right),
\\
{\cal V}
&=& - \sum_{i=1}^{N_S} q_i \delta {\cal F}_i(\{ q_{\alpha} \})
\end{eqnarray}
where $V(\{ q_i \})$ is the interaction potential 
function between $N_S$ system modes which can be described by, 
for example, the reduced model, Eq.~(\ref{eq:reduced}).
The simplest case $V(\{ q_i \})=0$ is trivial 
as each system mode may be treated seperately within 
the perturbation approximation for ${\cal V}$.

We assume that $|k \rangle$ is a certain state 
in the Hilbert space spanned by ${\cal H}_S$.
Then the reduced density matrix is 
\begin{eqnarray}
(\rho_S)_{mn}(t)
&=&
\langle m| e^{-i {\cal H}_S t/\hbar} {\rm Tr}_B \{ \tilde{\rho}(t) \} e^{i {\cal H}_S t/\hbar} |n \rangle
\end{eqnarray}
where tilde denotes the interaction picture.
Substituting the time-dependent perturbation expansion 
\begin{eqnarray}
\tilde{\rho}(t)
&=&
\rho(0)+ \frac{1}{i \hbar} \int_0^t dt' 
[\tilde{\cal V}(t'), \rho(0)]
\nonumber
\\
&&
+\frac{1}{(i \hbar)^2} \int_0^t dt' \int_0^{t'} dt''  
[\tilde{\cal V}(t'), [\tilde{\cal V}(t''),\rho(0)]]
+ \cdots
\end{eqnarray}
into the above, we find
\begin{eqnarray}
(\rho_S)_{mn}(t)
&\simeq&
(\rho_S)^{(0)}_{mn}(t)
+(\rho_S)^{(1)}_{mn}(t)
+(\rho_S)^{(2)}_{mn}(t)
+\cdots,
\label{eq:density}
\end{eqnarray}
where 
\begin{eqnarray}
(\rho_S)^{(0)}_{mn}(t)
&=&
\langle m| e^{-i {\cal H}_S t/\hbar} {\rho}_S(0) e^{i {\cal H}_S t/\hbar} |n \rangle,
\nonumber
\\
&=&\langle m(-t)|{\rho}_S(0)|n(-t) \rangle
=\langle m| {\rho}_S(t) |n \rangle
\label{eq:density1}
\\
(\rho_S)^{(2)}_{mn}(t)
&=&
\frac{1}{(i \hbar)^2}
\int^{t}_0 dt' 
\int^{t'}_0 dt'' 
\langle m| e^{-i {\cal H}_S t/\hbar} 
{\rm Tr}_B \{ [\tilde{\cal V}(t'),[\tilde{\cal V}(t''),\rho(0)]]
\} 
e^{i {\cal H}_S t/\hbar} | n \rangle 
\nonumber
\\
&=&
\frac{1}{(i \hbar)^2} \int_0^t dt' \int_0^{t'} dt''  
\sum_{i,j}
\langle m(-t)|
[q_i(t') q_j(t'') \rho_S(0) 
\nonumber 
\\
&&
-q_j(t'') \rho_S(0) q_i(t')]
| n(-t) \rangle 
\langle \delta {\cal F}_i(t') \delta {\cal F}_j(t'') \rangle_B
\nonumber
\\
&+&
\frac{1}{(i \hbar)^2} \int_0^t dt' \int_0^{t'} dt''  
\sum_{i,j}
\langle m(-t)|
[\rho_S(0) q_j(t'') q_i(t') 
\nonumber
\\
&&-q_i(t') \rho_S(0) q_j(t'')]
|n(-t) \rangle 
\langle \delta {\cal F}_j(t'') \delta {\cal F}_i(t') \rangle_B.
\label{eq:density2}
\end{eqnarray}
Here we have defined $|m(t) \rangle=e^{-i {\cal H}_S t/\hbar} | m \rangle$ and 
taken $(\rho_S)^{(1)}_{mn}(t)=0$.
Recognizing that we must evaluate expressions of the form
\begin{eqnarray}
R_{mn;ij}(t;t',t'')
&=&
\langle m(-t)|
[q_i(t') q_j(t'') \rho_S(0) 
| n(-t) \rangle,
\nonumber
\\
&&
-
\langle m(-t)|
q_j(t'') \rho_S(0) q_i(t')]
| n(-t) \rangle 
\\
C_{ij}(t',t'')
&=&
\langle \delta {\cal F}_i(t') \delta {\cal F}_j(t'') \rangle_B
\end{eqnarray}
and their complex conjugates, $R^*_{nm;ij}(t;t',t''), C^*_{ij}(t',t'')$,
the 2nd order contribution can be written
\begin{eqnarray}
(\rho_S)^{(2)}_{mn}(t)
&=&
\frac{1}{(i \hbar)^2}
\int^{t}_0 dt' 
\int^{t'}_0 dt'' 
\sum_{i,j} 
[R_{mn;ij}(t;t',t'') C_{ij}(t',t'')
\nonumber
\\
&&+R^*_{nm;ij}(t;t',t'') C^*_{ij}(t',t'')].
\end{eqnarray}
We can seperately treat the two terms. 
Assuming that we can  
solve ${\cal H}_S |a \rangle =E_a | a \rangle$, 
we find
\begin{eqnarray}
R_{mn;ij}(t;t',t'')
&=&
\sum_{abcd}
\langle m| a \rangle
(q_i)_{ab} (q_j)_{bc} (\rho_S)_{cd} 
\langle d| n \rangle  
\nonumber
\\
&&\times e^{-i(E_a-E_d)t-i(E_b-E_a)t'-i(E_c-E_b)t''}
\nonumber
\\
&-&
\sum_{abcd}
\langle m| a \rangle
(q_j)_{ab} (\rho_S)_{bc}  (q_i)_{cd} 
\langle d| n \rangle  
\nonumber
\\
&&\times
e^{-i(E_a-E_d)t-i(E_d-E_c)t'-i(E_b-E_a)t''}
\end{eqnarray}
For the bath-averaged term, we assume the following force 
due to third-order nonlinear coupling of system mode $i$
to the normal modes, $\alpha$ and $\beta$, of the bath
\cite{Fujisaki05}
\begin{equation}
\delta {\cal F}_i(\{ q_{\alpha} \})
=\sum_{\alpha,\beta} 
C_{i \alpha \beta}
(q_{\alpha} q_{\beta} - \langle q_{\alpha} q_{\beta} \rangle)
\label{eq:force}
\end{equation}
and we have \cite{Fujisaki05}
\begin{eqnarray}
C_{ij}(t',t'')
=R^{--}_{ij}(t',t'')+R^{++}_{ij}(t',t'')+R^{+-}_{ij}(t',t'')
\end{eqnarray}
with
\begin{eqnarray}
R^{--}_{ij}(t',t'')
&=&
\frac{\hbar^2}{2}
\sum_{\alpha,{\beta}} 
D_{\alpha {\beta};ij}
(1+n_{\alpha})(1+n_{{\beta}})
e^{-i(\omega_{\alpha} +\omega_{{\beta}})(t'-t'')
}, 
\\
R^{++}_{ij}(t',t'')
&=&
\frac{\hbar^2}{2}
\sum_{{\alpha},{\beta}} 
D_{{\alpha}{\beta};ij}
n_{\alpha} n_{{\beta}}
e^{i(\omega_{\alpha} +\omega_{{\beta}})(t'-t'')
}, 
\\
R^{+-}_{ij}(t',t'')
&=&
\hbar^2
\sum_{\alpha,{\beta}} 
D_{{\alpha}{\beta};ij}
(1+n_{\alpha})n_{\beta}
e^{-i(\omega_{\alpha} -\omega_{{\beta}})(t'-t'')
}
\end{eqnarray}
where
\begin{equation}
D_{\alpha\beta;ij}
=
\frac{C_{i \alpha \beta} C_{j \alpha \beta}} 
{\omega_{\alpha} \omega_{\beta}} 
\end{equation}
and $n_{\alpha}$ is the thermal population of the bath mode $\alpha$.

This formula reduces to our previous 
result for a one-dimensional system oscillator \cite{FZS06} when $N_S=1$ and 
all indices $(i,j)$ are suppressed. 
Importantly, this formula can be applied to situations 
where it is difficult to define a ``good'' normal mode to 
serve as a one-dimensional 
relaxing mode, as in the case of the CH stretching modes of 
a methyl group \cite{Fujisaki05}.
However, expanding to an $N_S$ dimensional system 
adds the burden of solving the multidimensional Schr\"odinger 
equation ${\cal H}_S |a \rangle =E_a | a \rangle$.
To address this challenge we may employ 
vibrational self-consistent field (VSCF) theory
and its extensions developed by Bowman and coworkers \cite{Bowman}
implemented in MULTIMODE program of Carter and Bowman \cite{MULTIMODE} 
or in the SINDO program \cite{sindo} of Yagi and coworkers.
As in the case of our previous theory of 
a one-dimensional system mode,
we must calculate $N_S$-tiple 3rd order coupling 
constants $C_{i \alpha \beta} (i=1,2, ... N_S)$ 
for all bath modes $\alpha$ and $\beta$.

\subsection{One-dimensional relaxing mode coupled to a fluctuating bath}
\label{sec:VER2}

We start from the following  time-dependent Hamiltonian
\begin{eqnarray}
{\cal H}(t)
&=& {\cal H}^0_S(t)+{\cal H}_B(t)+{\cal V}^0(t)
\\
&=&
{\cal H}^0_S(t)+\langle {\cal V}(t) \rangle_B
+{\cal H}_B(t)+{\cal V}^0(t)-\langle {\cal V}(t) \rangle_B
\\
&=&
{\cal H}_S(t)
+{\cal H}_B(t)+ {\cal V}(t)
\end{eqnarray}
where
\begin{eqnarray}
{\cal H}_S(t) 
&\equiv& {\cal H}^0_S(t)+\langle {\cal V}(t) \rangle_B,
\\
{\cal V}(t)
&\equiv&
{\cal V}^0(t)-\langle {\cal V}(t) \rangle_B
\end{eqnarray}
with the goal of solving the time-dependent Schr\"odinger equation
\begin{equation}
i \hbar \frac{\partial | \Psi(t) \rangle}{\partial t}
=[{\cal H}_S(t)
+{\cal H}_B(t)+  {\cal V}(t)]
| \Psi(t) \rangle
=[{\cal H}_0(t)
+{\cal V}(t)]
| \Psi(t) \rangle.
\end{equation}
By introducing a unitary operator $U_0(t)=U_S(t) U_B(t)$
\begin{eqnarray}
i \hbar \frac{d}{dt} U_0(t)
&=& {\cal H}_0(t) U_0(t),
\\
i \hbar \frac{d}{dt} U_S(t)
&=& {\cal H}_S(t) U_S(t),
\\
i \hbar \frac{d}{dt} U_B(t)
&=& {\cal H}_B(t) U_B(t),
\end{eqnarray}
we can derive an ``interaction picture'' von Neumann equation
\begin{equation}
i \hbar \frac{d}{dt} \tilde{\rho}(t)
= [\tilde{\cal V}(t),\tilde{\rho}(t)]
\end{equation}
where 
\begin{eqnarray}
\tilde{\cal V}(t)
&=& U_0^{\dagger}(t) {\cal V}(t) U_0(t),
\\
\tilde{\rho}(t)
&=& U_0^{\dagger}(t) \rho(t) U_0(t).
\end{eqnarray}
We assume the simple form of a 
harmonic system and bath, but allow fluctuations in the 
system and bath modes modeled by time-dependent frequencies 
\begin{eqnarray}
{\cal H}_S(t) &=& \hbar \omega_S(t) (a_S^{\dagger} a_S+1/2),
\\
{\cal H}_B(t) &=& \sum_{\alpha} 
\hbar \omega_{\alpha}(t) (a_{\alpha}^{\dagger} a_{\alpha}+1/2).
\end{eqnarray}
The unitary operators generated by these Hamiltonians are 
\begin{eqnarray}
U_S(t) &=& e^{-i \int_0^t d \tau \omega_S(\tau) (a_S^{\dagger} a_S+1/2)},
\\
U_B(t) &=& e^{-i \int_0^t d \tau \sum_{\alpha} \omega_{\alpha}(\tau) .
(a_{\alpha}^{\dagger} a_{\alpha}+1/2)}
\end{eqnarray}
and the time evolution of the annihilation operators 
is given by
\begin{eqnarray}
U_S^{\dagger}(t)a_S U_S(t)
&=& a_S e^{-i \int_0^t d \tau \omega_S(\tau)},
\\
U_{B}^{\dagger}(t)a_{\alpha} U_{B}(t)
&=& a_{\alpha} e^{-i \int_0^t d \tau \omega_{\alpha}(\tau)}.
\end{eqnarray}
To simplify the evaluation of the force autocorrelation 
function, we assume that the temperature is low or the system mode 
frequency is high as a justification for the 
approximation. 
Substituting the above result into the force autocorrelation
function calculated by the force operator, Eq.~(\ref{eq:force}),
we find 
\begin{eqnarray}
\langle \delta {\cal F}(t') \delta {\cal F}(t'') \rangle 
&\simeq& 
\frac{\hbar^2}{2}
\sum_{\alpha,{\beta}} 
\frac{C_{S \alpha \beta}(t') C_{S \alpha \beta}(t'')} 
{\sqrt{\omega_{\alpha}(t') \omega_{\beta}(t') \omega_{\alpha}(t'') \omega_{\beta}(t'')}} 
\nonumber
\\
&&\times
e^{-i [\Theta_{\alpha \beta}(t') -\Theta_{\alpha \beta}(t'')]} 
\end{eqnarray}
where 
\begin{eqnarray}
\Theta_{S}(t) 
&=&  \int_0^t d \tau \omega_S(\tau),
\\
\Theta_{\alpha \beta}(t)
&=&  \int_0^t d \tau [\omega_{\alpha}(\tau)+\omega_{\beta}(\tau)].
\end{eqnarray}
Substituting this approximation into the 
perturbation expansion Eqs.~(\ref{eq:density}), (\ref{eq:density1}), 
and (\ref{eq:density2}),
we obtain our final result
\begin{eqnarray}
(\rho_S)_{00}(t)
&\simeq& 
\frac{\hbar}{ 2}
\sum_{\alpha,\beta}
\int_0^t dt' \int_0^{t'} dt''  
\frac{C_{S \alpha \beta}(t') C_{S \alpha \beta}(t'')} 
{\sqrt{\omega_S(t') \omega_{\alpha}(t') \omega_{\beta}(t') \omega_S(t'') \omega_{\alpha}(t'') \omega_{\beta}(t'')}} 
\nonumber
\\
&&
\times
\cos 
\left \{
\Theta_S(t')-\Theta_{\alpha \beta}(t') - 
\Theta_S(t'')+\Theta_{\alpha \beta}(t'') 
\right \}
\label{eq:VER-time}
\end{eqnarray}
which provides a dynamic 
correction to the previous formula \cite{FZS06}.
The time-dependent parameters 
$\omega_S(t), \omega_{\alpha}(t), C_{S \alpha \beta}(t)$ 
may be computed from 
a running trajectory using instantaneous normal mode analysis \cite{Keyes}.
This result was first derived by Fujisaki and Stock \cite{FS08},
and applied to the VER dynamics of $N$-methylacetamide as described below.
This correction eliminates the assumption 
that the bath frequencies are static on 
the VER timescale.

For the case of a static bath,
the frequency and coupling parameters 
are time-independent and this formula reduces to the 
previous one-dimensional formula (when the 
off-resonant terms are neglected) \cite{FZS06}:
\begin{eqnarray}
(\rho_S)_{00}(t)
&\simeq& 
\frac{\hbar}{ 2 \omega_S}
\sum_{\alpha,\beta}
\frac{C_{S \alpha \beta}^2} 
{\omega_{\alpha} \omega_{\beta}} 
\frac{1-\cos [ (\omega_S-\omega_{\alpha}-\omega_{\beta})t ]} 
{(\omega_S-\omega_{\alpha}-\omega_{\beta})^2}.
\label{eq:VER-1d}
\end{eqnarray}

Note that Bakker derived a similar fluctuating Landau-Teller formula
 in a different manner \cite{Bakker04}. It was successfully applied to 
molecular systems by Sibert and coworkers \cite{Sibert}. 
However, the above formula differs from Bakker's as
(a) we use the instantaneous normal mode analysis to parameterize 
our expression,
and (b) we do not take the Markov limit. 
Our formula 
can describe the time course of the density matrix as well 
as the VER rate.

One further point is that we use the cumulant-type approximation 
to calculate the dynamics.
When we calculate an excited state probability,
we use 
\begin{equation}
(\rho_S)_{11}(t)=1-(\rho_S)_{00}(t) \simeq \exp 
\{ -(\rho_S)_{00}(t) \}.
\label{eq:cumulant}
\end{equation}
Of course, this is valid for the initial process 
($(\rho_S)_{00}(t) \ll 1$),
but, at longer time scales, 
we take $(\rho_S)_{11}(t) \simeq \exp 
\{ -(\rho_S)_{00}(t) \}$ because the naive 
formula $(\rho_S)_{11}(t)=1-(\rho_S)_{00}(t)$ can be 
negative, which is unphysical \cite{FS08}.

\subsection{Limitations of the VER formulas and comments}

There are several limitations to the VER formulas derived above.
The most obvious is that they are 2nd order perturbative formulas 
and rely on a short-time approximation. As far as we know, however, 
there exists no non-perturbative quantum mechanical treatment 
of VER applicable to {\it large} molecular systems.
It is prohibitive to treat the full molecular dynamics
quantum mechanically  \cite{GW04} for large molecules.
Moreover, while there exist several mixed quantum-classical 
methods \cite{Okazaki} that may be applied to the study of VER,
but there is no guarantee that such approximate methods work 
better than the perturbative treatment \cite{BB94}.

Another important limitation is the adaptation of a 
normal mode basis set, a natural choice for molecular vibrations.
Because of the normal mode analysis, the computation 
can be burdensome. When we employ instantaneous normal mode
analysis \cite{Keyes}, there is a concern about the 
imaginary frequency modes. For the study of high 
frequency modes, this may not be significant. However, 
for the study of low frequency modes, the divergence
of quantum (or classical) dynamics due to the presence 
of such imaginary frequency modes is a significant concern.
For the study of low frequency modes, 
it is more satisfactory to use other methods that do not rely 
on normal mode analysis such as semiclassical methods \cite{Geva}
or path integral methods \cite{pathintegral}. 

We often use ``empirical'' force fields to describe quantum 
dynamics. However, it is well known that the 
force fields underestimate anharmonicity of molecular 
vibrations \cite{GCG02}. It is often desirable to use {\it ab initio} 
potential energy surfaces. 
However, that more rigorous approach can be demanding.
Lower levels of theory can fail to match the 
accuracy of some empirical potentials.
As a compromise, 
approximate potentials of intermediate accuracy, 
such as QM/MM potentials \cite{SF09}, may be appropriate.
We discuss this question further in Sec.~\ref{sec:NMA} and 
in Sec.~\ref{sec:porphyrin}.

\section{Applications of the VER formulas to 
vibrational modes in biomolecules}
\label{sec:application}

We study the quantum dynamics of 
high frequency modes in biomolecular systems 
using a variety of VER formulas described in Sec.~\ref{sec:derivation}.
The application of a variety of theoretical approaches 
to essential VER processes will allow for a relative 
comparison of theories as well as the absolute assessment 
of theoretical predictions compared with experimental 
observations. In doing so, we address a number of 
fundamental questions.
What are the limitations of the static bath approximation for 
fast VER in biomolecular systems?
Can the relaxation dynamics of a relaxing amide I vibration
in a protein be accurately modeled as a one-dimensional
system mode coupled to a harmonic bath?
Can the ``fluctuating bath'' model accurately capture 
the system dynamics when the static picture of normal modes is 
not ``good'' on the timescale of the VER process?
In Sec.~\ref{sec:NMA} and \ref{sec:cytc}, 
our main focus is the VER of excited 
amide I modes in peptides or proteins. 
In Sec.~\ref{sec:porphyrin}, 
we study some vibrational modes in 
porphyrin ligated to imizadol, which 
is a mimic of a heme molecule in heme-proteins including myoglobin and hemoglobin.

\subsection{$N$-methylacetamide}
\label{sec:NMA}

$N$-methylacetamide (NMA) is a well-studied small 
molecule (CH$_3$-CO-NH-CH$_3$) that serves as a convenient model of a peptide bond
structure (-CO-NH-) in theory and experiment. 
As in other amino acids, 
there is an amide I mode,
localized on the CO bond stretch, 
which is a useful ``reporter'' of peptide structure and dynamics when probed by 
infrared spectrocopy. 
Many theoretical and experimental studies on amide I and
other vibrational modes (amide II, III) have 
characterized how the mode frequencies depend on 
the local secondary structure of 
peptides or proteins \cite{Krimm,TT92}.
For the accurate description of frequencies and polarizability of these modes,
see Refs.~\cite{Knoester07,Knoester08,TT98,Skinnergroup,HHC05,Mukamelgroup,
Stock06}.
The main focus of these works is the frequency sensitivity of amide 
modes on the molecular configuration and environment. 
In this case, the amide mode frequencies are treated in a quantum 
mechanical way, but the configuration is treated classically.
With a focus on interpreting mode frequency shifts due to 
configuration and environment,
mode-coupling between amide modes and other 
modes is often neglected. 
As we are mainly interested in VER or IVR dynamics of 
these modes, an accurate treatment of the mode coupling 
is essential.

\begin{figure}[b]
\hfill
\begin{center}
\includegraphics[scale=0.7]{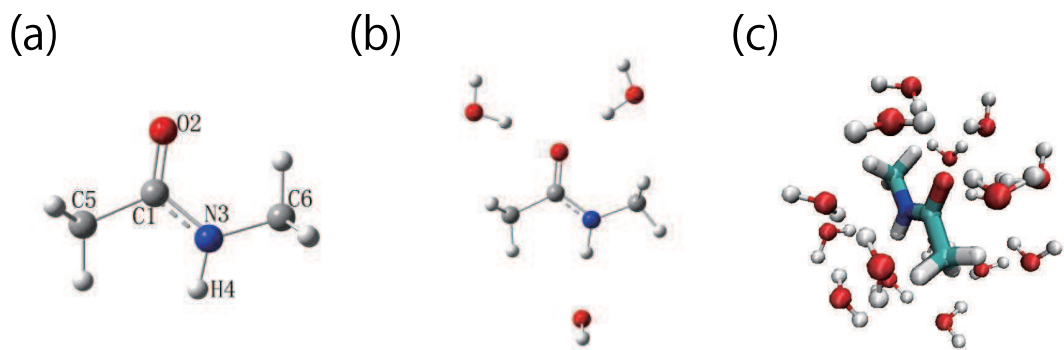}
\end{center}
\caption{\baselineskip5mm
Representation of three models 
employed for the study of VER dynamics in $N$-methylacetamide (NMA).
(a) NMA, (b) NMA with three solvating water, 
(c) NMA with first solvation shell derived from 
simulations in bulk water.
[Reprinted with permission for (a) and (b) from 
Y. Zhang, H. Fujisaki, and J.E. Straub,
J.~Phys.~Chem.~A {\bf 113}, 3051 (2009).
Copyright \copyright 2009 by the American Chemical Society.]
[Reprinted with permission for (c) from 
H. Fujisaki and G. Stock,
J.~Chem.~Phys.~{\bf 129},134110 (2008).
Copyright \copyright 2009 by the American Institute of Physics.]
}
\label{fig:nma}
\end{figure}

Recent theoretical development of IVR dynamics 
in small molecules is summarized in \cite{GW04}. 
Leitner and Wolynes \cite{LW97} utilized the concept of 
local random matrix to clarify the quantum aspects 
of such dynamics. The usefulness and applications of their 
approach are summarized in \cite{Leitner} 
as well as in this volume \cite{Leitner09}.
However, these studies are focused on isolated molecules, 
whereas our main interest is in exploring quantum dynamics 
in a condensed phase.
We take a step-by-step hierarchical approach.
Starting from the isolated NMA molecule, we add several water 
molecules to form NMA-water clusters, and finally treat the
condensed phase NMA-water system (see Fig~\ref{fig:nma}).
With increasing complexity of our dynamical model,
the accuracy of our theory, including the quality of 
the potential energy surface and the accuracy 
of the quantum dynamics must diminish.
As such, a pricipal focus of our account 
is a careful examination and validation of 
our procedures through comparison with 
accurate methods or experiment.

\subsubsection{$N$-methylacetamide in vacuum}

\begin{figure}[b]
\hfill
\begin{center}
\begin{minipage}{.42\linewidth}
\includegraphics[scale=1.1]{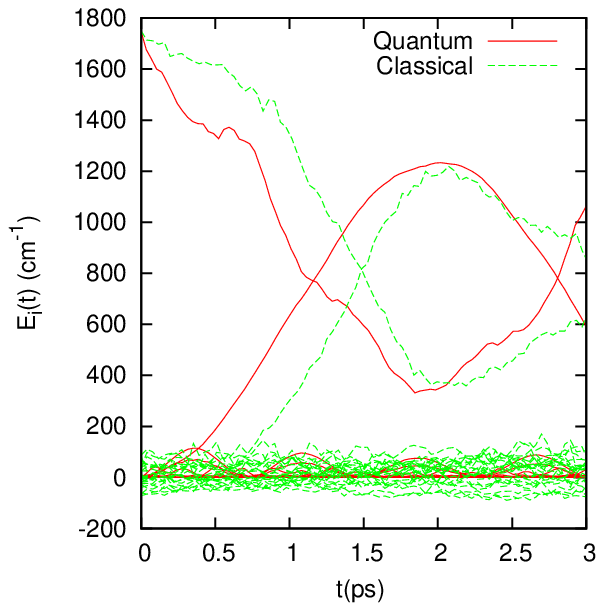}
\end{minipage}
\hspace{1cm}
\begin{minipage}{.42\linewidth}
\includegraphics[scale=1.1]{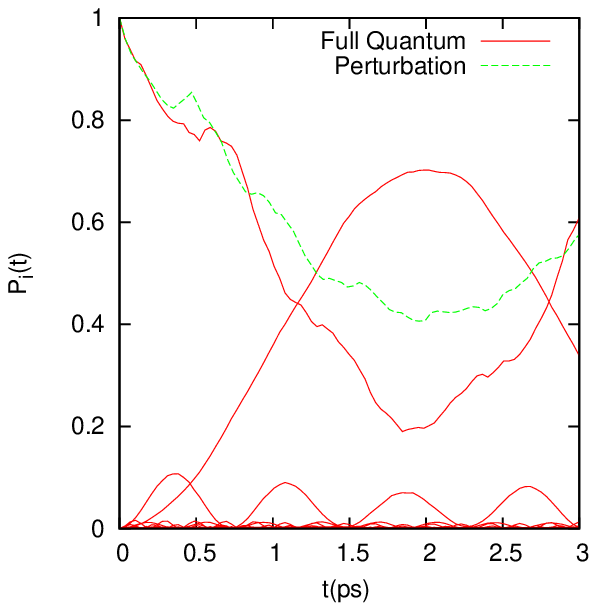}
\end{minipage}
\end{center}
\caption{\baselineskip5mm
Left: Time evolution of the energy content of the
initially excited amide I mode as well as all the remaining modes
of $N$-methylacetamide. Quantum (red lines) and classical (green
lines) calculations obtained at the DFT/B3LYP level of theory 
are compared. 
Right: Comparison of the VCI calculation (red) with the result of 
the perturbative calculation (green)
for the reduced density matrix. 
[Reprinted with permission from 
H. Fujisaki, K. Yagi, J.E. Straub, and G. Stock,
Int.~J.~Quant.~Chem. {\bf 109}, 2047 (2009).
Copyright \copyright 2009 by Wiley InterScience.]
}
\label{fig:vac1}
\end{figure}

\begin{figure}[b]
\hfill
\begin{center}
\includegraphics[scale=0.9]{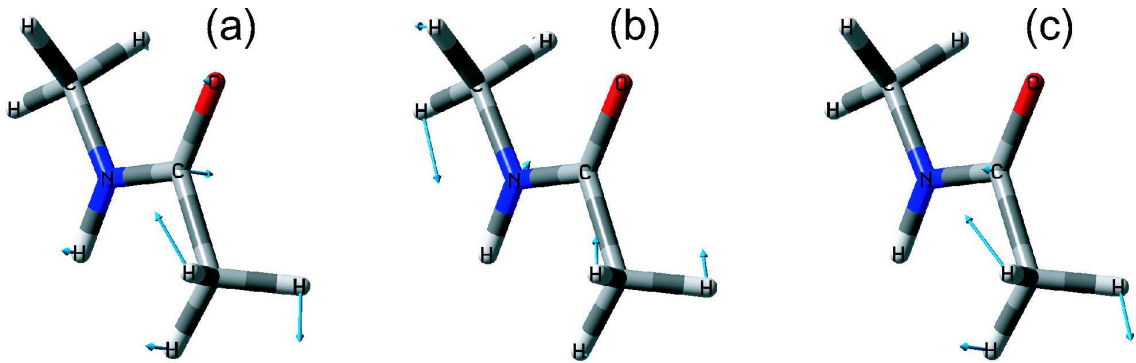}
\end{center}
\caption{\baselineskip5mm
The dominant bath vibrational modes coupled to the 
amide I mode calculated on the B3LYP/6-31G(d) potential
energy surface. 
[Reprinted with permission from 
H.~Fujisaki, K.~Yagi, K.~Hirao, and J.E. Straub,
Chem. Phys. Lett. {\bf 443}, 6 (2007).
Copyright \copyright 2009 by Elsevier.]
}
\label{fig:path}
\end{figure}

In our studies of isolated NMA \cite{FYHS07,FYSS09},
we have employed both
accurate potential energy surface 
and accurate quantum dynamics methods to explore the timescale 
and mechanism of VER.
From the anharmonic frequency calculations and comparison to 
experiment \cite{ATT84}, 
we concluded that B3LYP/6-31G(d) is 
a method of choice for computation of the electronic ground state 
potential surface, 
considering both accuracy and feasibility.
For other treatments at differing levels of theory 
of quantum chemical calculation on NMA,
see Refs.~\cite{GCG02,BS06,KB07}.
After the construction of an accurate potential surface, 
there are several
tractable approaches for treating the quantum dynamics for this system.
The most accurate is the vibrational configuration interaction (VCI) method
based on vibrational self-consistent field (VSCF) basis sets 
(see Refs.~\cite{Bowman,MULTIMODE,FYHS07,FYSS09} for details).
We employed the Sindo code developed by Yagi \cite{sindo}.
The numerical results for the VCI calculation 
are shown in Fig.~\ref{fig:vac1},
and compared to the prediction based on the perturbative formula Eq.~(\ref{eq:VER-1d})
and classical calculations as done in \cite{Stock}.
Both approximate methods seem to work well, but 
there are caveats. 
The perturbative formula only works at short time scales.
There is ambiguity for the classical simulation 
regarding how the zero point energy correction should be 
included (see Stock's papers \cite{Stock99}).
The main results for a singly deuterated NMA (NMA-$d_1$) 
are (1) the relaxation time appears to be sub ps,
(2) as NMA is a small molecule, there is a recurrent phenomenon,
(3) the dominant relaxation pathway involves three bath modes as shown in Fig.~\ref{fig:path},
and (4) the dominant pathways can be identified and 
characterized by the following 
Fermi resonance parameter \cite{FYHS07,FYSS09}
\begin{equation}
\eta 
\equiv
\left| 
\frac{\langle i| \Delta V|f \rangle}{\Delta E}
\right|
\propto
\left| 
\frac{C_{Skl}}
{\hbar(\omega_S - \omega_k-\omega_l)}
\right|
\sqrt{\frac{\hbar}{2 \omega_S}}
\sqrt{\frac{\hbar}{2 \omega_k}}
\sqrt{\frac{\hbar}{2 \omega_l}}
\label{eq:Fermi}
\end{equation}
where $\langle i| \Delta V|f \rangle$ is the matrix element for the 
anharmonic coupling interaction, 
and $\Delta E=\hbar(\omega_S - \omega_k-\omega_l)$
is the resonance condition (frequency matching) 
for the system and two bath modes.
Both the resonant condition ($\Delta E$) and anharmonic coupling elements 
($C_{Skl}$) play a role, but we found that 
the former affects the result more significantly.
This indiates that, for the description of VER phenomena in molecules,
accurate calculation of the harmonic 
frequencies is more important than the accurate calculation of anharmonic 
coupling elements. This observation is the basis for the 
development and application of the 
multi-resolution methods for anharmonic frequency calculations \cite{Yagi,Raufut}.

\subsubsection{$N$-methylacetamide/Water cluster}

\begin{figure}[b]
\hfill
\begin{center}
\includegraphics[scale=1.2]{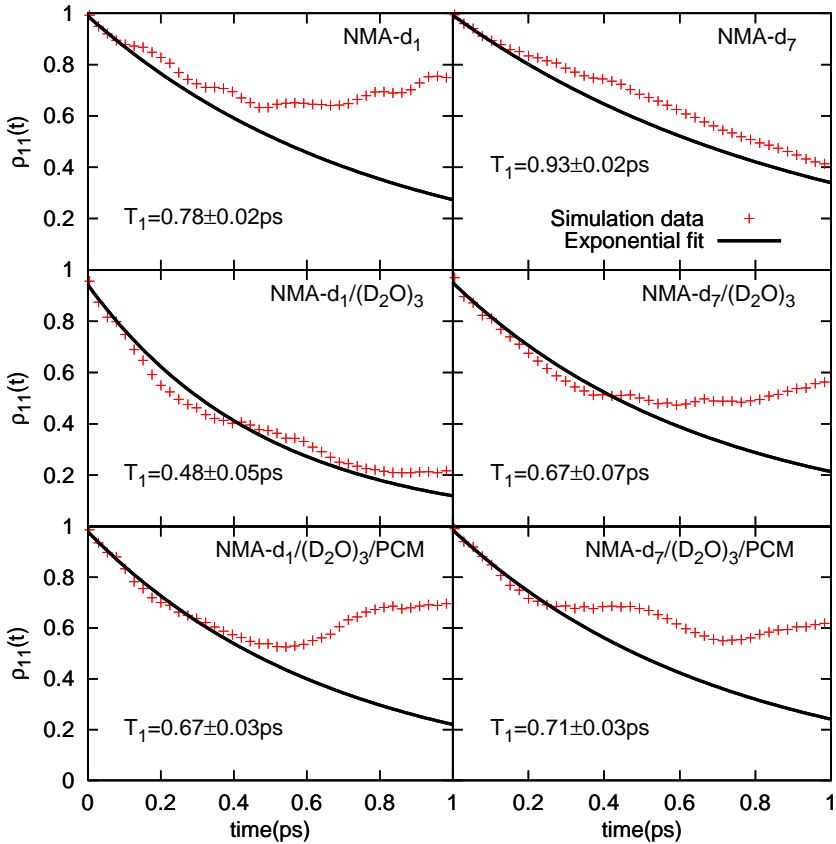}
\end{center}
\caption{\baselineskip5mm
Time evolution of the density matrix for the 
amide I mode in the NMA-water cluster system 
after $v=1$ excitation. 
The derived vibrational energy relaxation 
time constants $T_1$ are also provided.
[Reprinted with permission from 
Y. Zhang, H. Fujisaki, and J.E. Straub,
J.~Phys.~Chem.~A {\bf 113}, 3051 (2009).
Copyright \copyright 2009 by the American Chemical Society.]
}
\label{fig:nma-water}
\end{figure}

We next examine a somewhat larger system, NMA in a water 
cluster \cite{ZFS09b},
an interesting and important model system for exploring the 
response of amide vibrational modes to ``solvation'' \cite{Skinnergroup}.
The system size allows for an ab initio quantum mechanical treatment of the potential surface at a 
higher level of theory, B3LYP/aug-cc-pvdz, relative to the commonly employed B3LYP/6-31G(d).
The enhancement in level of theory significantly improves the quality of the
NMA-water interaction,
 specifically the structure and energetics of hydrogen bonding.
Since there are at most three hydrogen bonding sites in NMA, 
it is natural to configure three water molecules around NMA as a minimal model of ``full solvation.''
NMA-water hydrogen bonding causes the frequency of the amide I mode to
red-shift. 
As a result, the anharmonic coupling between the relaxing 
mode and the other bath modes will change 
relative to the case of the isolated NMA.
Nevertheless, we observe that the VER time scale remains sub ps as is the 
case for isolated NMA (Fig.~\ref{fig:nma-water}).
Though there are intermolecular (NMA-water) contributions 
to VER, they do not significantly alter the VER timescale.
Another important finding is that the energy pathway from the 
amide I to amide II mode is ``open'' for the NMA-water cluster system.
This result is in agreement with experimental results by 
Tokmakoff and coworkers \cite{Tokmakoff06} and 
recent theoretical investigation \cite{Knoester08}.
Comparison between singly (NMA-$d_1$) and fully (NMA-$d_7$) 
deuterated cases shows the VER time scale become somewhat longer 
for the case of NMA-$d_7$ (Fig.~\ref{fig:nma-water}). 
We also discuss this phenomenon below in the context of the NMA/solvent water 
system.

\subsubsection{$N$-methylacetamide in water solvent}
Finally we consider the condensed phase system of NMA
in bulk water \cite{FZS06,FS08,SF09}. We attempt to include 
the full dynamic effect of the system by generating 
many configurations from molecular dynamics 
simulations and using them to ensemble-average the results.
Note that in the previous examples of isolated NMA and 
NMA/water clusters, only one configuration at a local minimum of the
zero temperature ground state potential surface was used.
On the other hand, the potential energy function 
used is not so accurate as in the previous examples
as it is not feasible to include many 
water molecules at a high level of theory. 
We have used the CHARMM force field
to calculate the potential energy and to carry out 
molecular dynamics simulations.

All simulations were performed using the CHARMM simulation program
package \cite{CHARMM} and the CHARMM22 all-atom force
field \cite{CHARMMFF} was employed to model the solute 
NMA-$d_1$
and the
TIP3P water model \cite{TIP3P} with doubled hydrogen masses to model the
solvent D$_2$O. We also performed simulations for fully deuterated 
NMA-$d_7$.
The peptide was placed in a periodic cubic box
of (25.5 \AA)$^3$ containing 551 D$_2$O molecules. All bonds
containing hydrogens were constrained using the SHAKE 
algorithm \cite{SHAKE}.
We used a 10 \AA \, cutoff with a switching function for the nonbonded
interaction calculations. After a standard equilibration protocol, we
ran a 100 ps NVT trajectory at 300 K, from which 100 statistically
independent configurations were collected. 

\begin{figure}[b]
\hfill
\begin{center}
\begin{minipage}{.42\linewidth}
\includegraphics[scale=1.1]{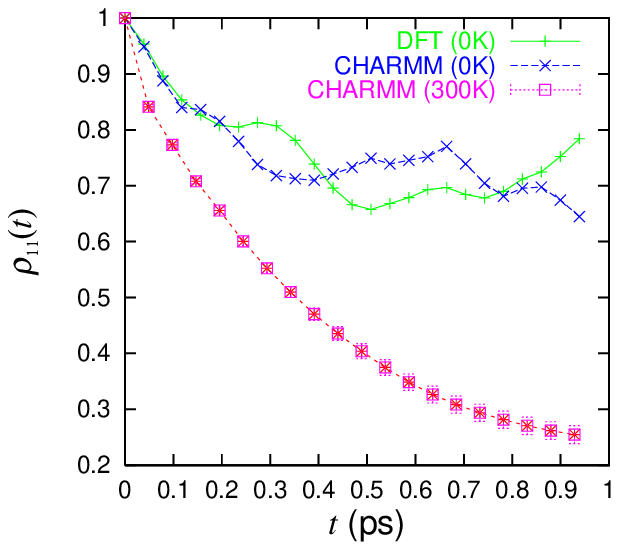}
\end{minipage}
\hspace{1cm}
\begin{minipage}{.42\linewidth}
\includegraphics[scale=1.1]{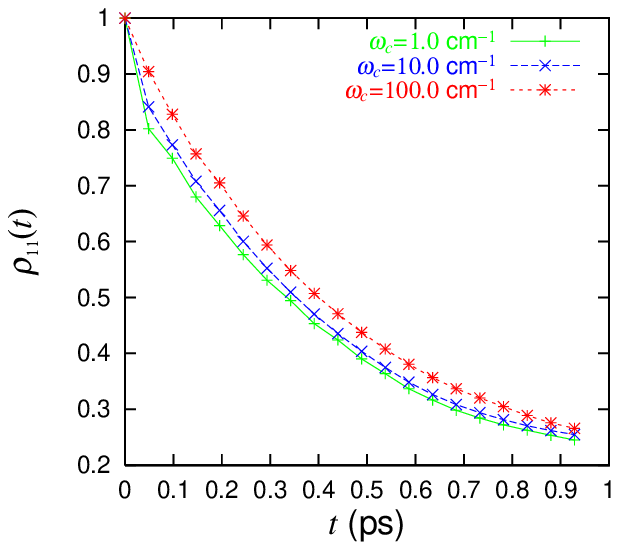}
\end{minipage}
\end{center}
\caption{\baselineskip5mm
Left: Comparison of the calculation of the $\rho_{11}$ element of the reduced density matrix at different
levels of theory.
Right: Calculation of the density matrix with different cutoff frequencies.
[Reprinted with permission from 
H. Fujisaki, Y. Zhang, and J.E. Straub,
J.~Chem.~Phys.~{\bf 124}, 144910 (2006).
Copyright \copyright 2009 by the American Institute of Physics.]
}
\label{fig:VER-FZS06}
\end{figure}

We applied the simplest VER formula Eq.~(\ref{eq:VER-1d})~\cite{FZS06}
as shown in Fig.~\ref{fig:VER-FZS06}.
We truncated the system including only NMA 
and several water molecules around NMA with a cutoff distance,
taken to be 10 \AA.
For reasons of computational feasibility,
we only calculated the normal modes and anharmonic coupling 
elements within this subsystem.

A number of important conclusions were drawn from these calculations.
\\
(1) The inclusion of ``many'' solvating water molecules induces the irreversible 
decay of the excess energy as well as the density matrix elements (population).
The important observation is that the VER behavior does not 
severely depend on the cutoff distance (if it is large enough) and 
the cutoff frequency. 
The implication is that if we are interested in a localized mode 
such as the amide I mode in NMA, it is enough to use a NMA/water 
cluster system to totally describe the initial process of VER.
In a subsequent study, Fujisaki and Stock 
used only 16 water molecules surrounding NMA (hydrated water) and found
reasonable results \cite{FS08}.
\\
(2) Comparison of the two isolated NMA calculations suggests that 
the CHARMM force field works well compared with results based on DFT calculations.
This suggests that the use of the empirical force field in exploring
VER of the amide I mode may be justified.
\\
(3) There is a classical limit of this calculation \cite{FZS06},
which predicts a slower VER rate close to 
Nguyen-Stock's quasi-classical calculation \cite{Stock}. 
This finding was explored further by Stock \cite{Stock09},
who derived a novel quantum correction factor 
based on the reduced model, Eq.~(\ref{eq:reduced}).

\begin{figure}[b]
\hfill
\begin{center}
\begin{minipage}{.42\linewidth}
\includegraphics[scale=1.1]{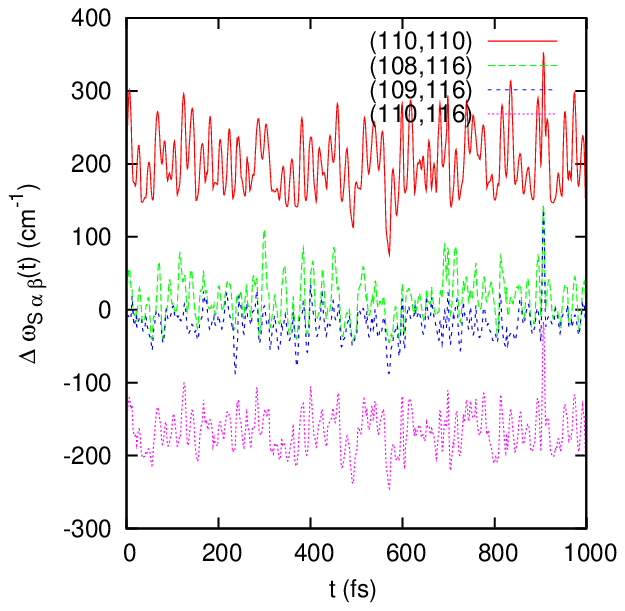}
\end{minipage}
\begin{minipage}{.42\linewidth}
\includegraphics[scale=1.1]{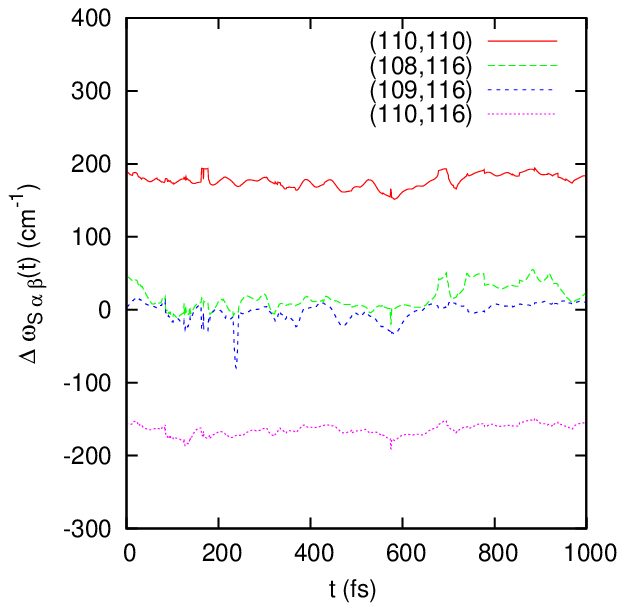}
\end{minipage}
\begin{minipage}{.42\linewidth}
\includegraphics[scale=1.1]{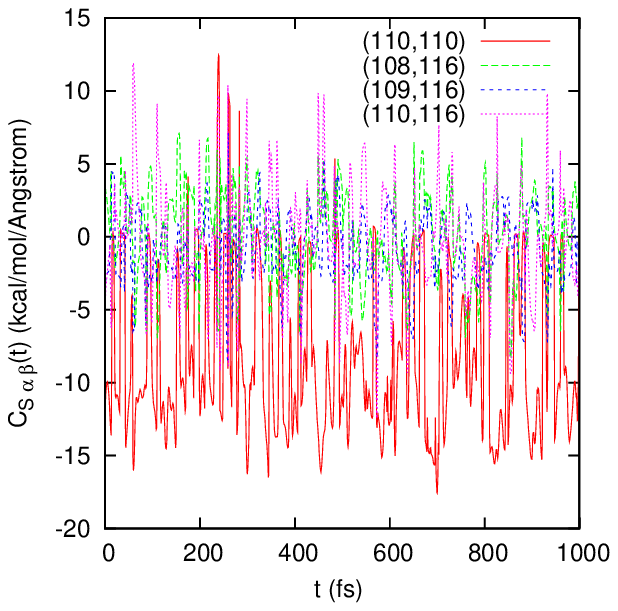}
\end{minipage}
\begin{minipage}{.42\linewidth}
\includegraphics[scale=1.1]{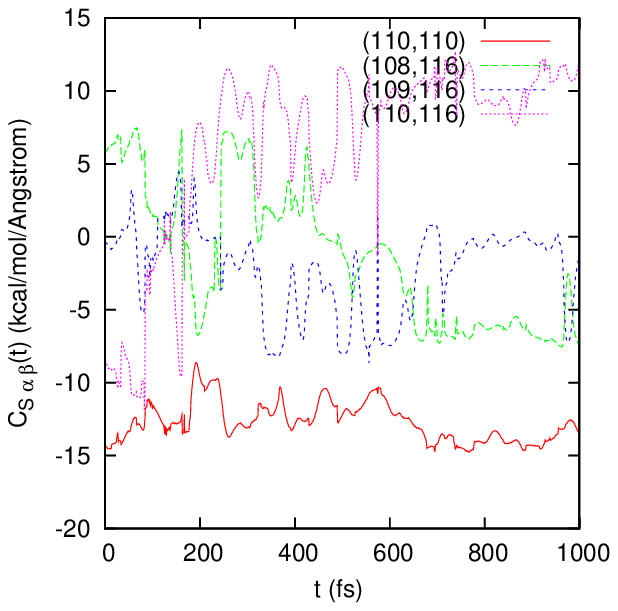}
\end{minipage}
\end{center}
\caption{\baselineskip5mm
Time evolution of the vibrational dynamics of
NMA in D$_2$O obtained from instantaneous
normal mode analysis with (right) and without (left) partial energy
minimization. Shown are (upper panels) 
the frequency mismatch $\Delta \omega_{S \alpha \beta}(t) = \omega_S(t)
-\omega_{\alpha}(t) - \omega_{\beta}(t)$, for several resonant bath
mode combinations, and (lower panels) the corresponding third-order
anharmonic couplings, $C_{S \alpha \beta}(t)$.
[Reprinted with permission from 
H. Fujisaki and G. Stock,
J.~Chem.~Phys.~{\bf 129},134110 (2008).
Copyright \copyright 2009 by the American Institute of Physics.]
}
\label{fig:amideI}
\end{figure}

In these calculations, many solvating water
configurations were generated using MD simulations.
As such,
information characterizing dynamic fluctuation in the environment is ignored.
Fujisaki and Stock further improved the methodology to 
calculate VER \cite{FS08} by taking into account the dynamic 
effects of the environment through the incorporation of time-dependent
parameters, such as the normal mode frequencies and 
anharmonic coupling,
derived from the MD simulations as shown in Fig.~\ref{fig:amideI}.
Their method is described in 
Sec.~\ref{sec:VER2}, and was applied to the same NMA/solvent water system.
As we are principally concerned with high frequency modes,
and the instantaneous normal mode frequencies 
can become unphysical, 
we adopted a partial optimization strategy.
We optimized the NMA under the influence of the 
solvent water at a {\it fixed} position.
(For a different strategy, see Ref.~\cite{Yagi09}.)
The right side of Fig.~\ref{fig:amideI} shows the numerical result 
of the optimization procedure.
Through partial optimization, 
the fluctuations of the parameters become milder 
than the previous calculations that employed instantaneous 
normal modes.

\begin{figure}[b]
\hfill
\begin{center}
\begin{minipage}{.42\linewidth}
\includegraphics[scale=1.1]{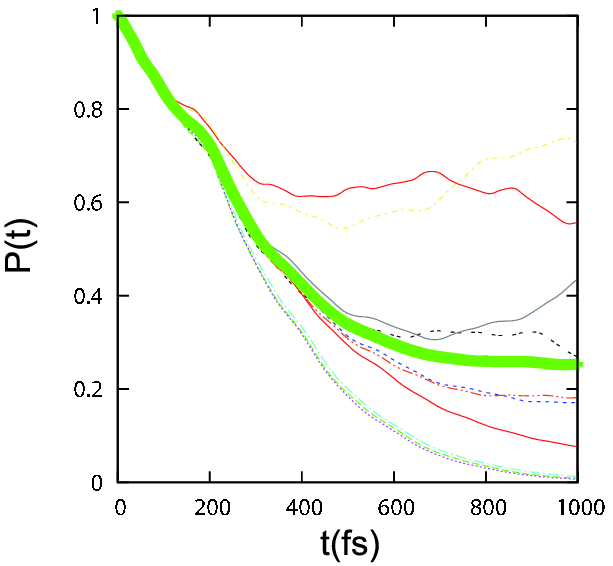}
\end{minipage}
\hspace{1cm}
\begin{minipage}{.42\linewidth}
\includegraphics[scale=1.1]{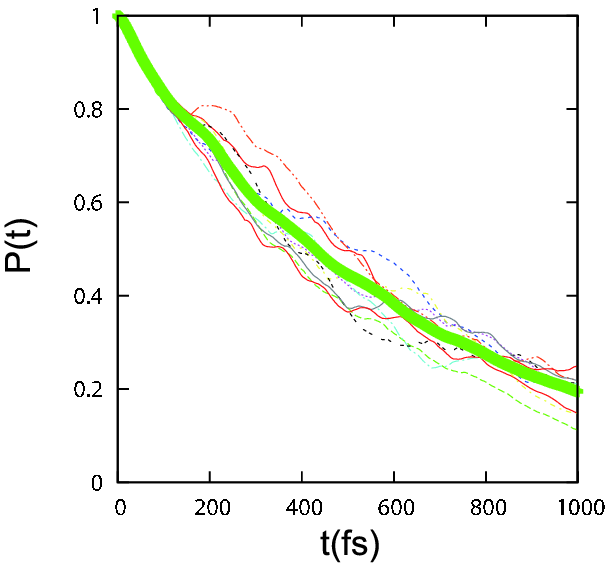}
\end{minipage}
\end{center}
\caption{\baselineskip5mm
VER calculations of amide I mode population $P(t)$ of NMA
with use of instantaneous normal mode analysis and
partial energy minimization. Shown are results from
(left) the inhomogeneous averaging approximation and (right) dynamical averaging.  
[Reprinted with permission from 
H. Fujisaki and G. Stock,
J.~Chem.~Phys.~{\bf 129},134110 (2008).
Copyright \copyright 2009 by the American Institute of Physics.]
}
\label{fig:VER-FS08}
\end{figure}

The results of numerical calculations based on Eq.~(\ref{eq:VER-time}) are 
shown in Fig.~\ref{fig:VER-FS08}. We see that  
both partial optimization and dynamical 
averaging affect the result.
The ``dynamic'' formula, Eq.~(\ref{eq:VER-time}), leads to smaller fluctuations in the 
results for the density matrix. 
Apparently, dynamic averaging smoothens the resonant effect,
stemming from the frequency difference in the denominator 
of Eq.~(\ref{eq:VER-1d}). For the NMA/solvent water system,
the time-averaged value of the Fermi resonance 
parameter, Eq.~(\ref{eq:Fermi}),  
can be utilized to clarify the VER pathways as in the case of
isolated NMA \cite{FS08}. It was shown that the 
hydrated water (the number of waters is 16) 
is enough to fully describe 
the VER process at the initial stage ($\simeq 0.5$ ps).
The predictions of the VER rates for the two deuterated cases, NMA-$d_1$ and NMA-$d_7$, 
are in good agreement with experiment and  also with the NMA/Water cluster 
calculations \cite{ZFS09b}.
Though the dynamic effect is modest in the case of the NMA/solvent water system, 
the dynamic formula is recommended when variations 
of the system parameters due to the fluctuating
environment must be taken into account.

\subsection{Cytochrome c in water}
\label{sec:cytc}

The protein, cytochrome c, has been used in experimental and theoretical studies of VER
\cite{Fujisaki05,Straub.JPCB.2003.107.12339,
Straub.JPCB.2009.113.825,Champion.JCP.1992.97.3214,
Champion.JPCB.2000.104.10789,Kruglik.JPCB.2006.110.12766,BS03}.
Importantly, spectroscopy and simulation have been used to explore the time scales and mechanism of VER of
CH stretching modes \cite{Fujisaki05,BS03}.
Here we examine VER of amide I modes in cytochrome c \cite{FS07}.
Distinct from previous studies \cite{Fujisaki05} which 
(a) employed a static local minimum 
of the system, we use the dynamical trajectory;
(b) in the previous study, the water degrees of freedom were   
excluded, whereas in this study some hydrating water has been 
taken into account.

\begin{figure}[b]
\hfill
\begin{center}
\begin{minipage}{.42\linewidth}
\includegraphics[scale=0.4]{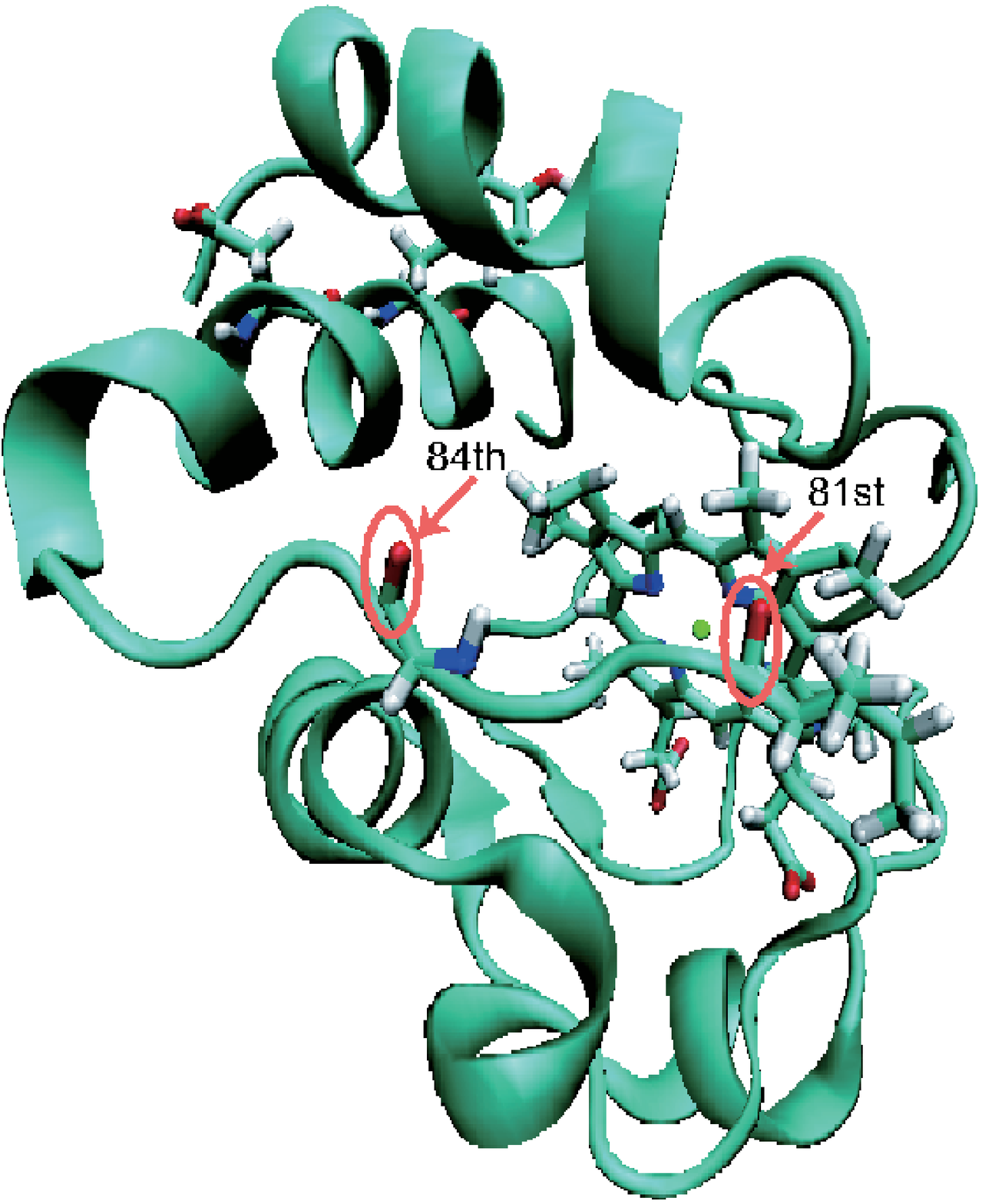}
\end{minipage}
\hspace{1cm}
\begin{minipage}{.42\linewidth}
\includegraphics[scale=0.4]{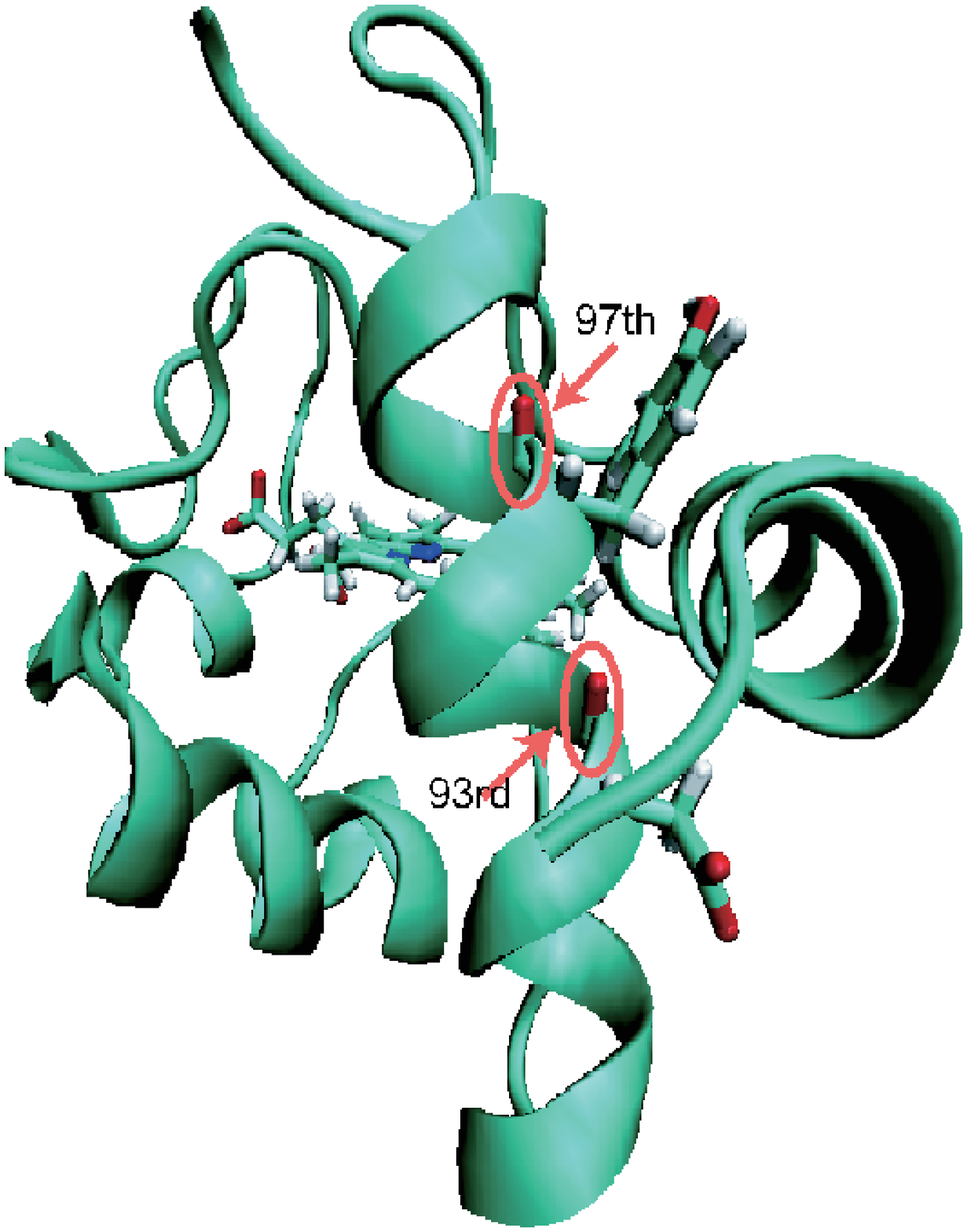}
\end{minipage}
\end{center}
\caption{\baselineskip5mm
Left: 
81st and 84th residues of cytochrome c in a loop region. 
Right: 
93rd and 97th residues of cytochrome c in an $\alpha$ helical region. 
The cartoon represents the protein
using a licorice model to identify the four residues.
(The water molecules are excluded for simplicity.)
[Reprinted with permission from 
H. Fujisaki and J.E. Straub,
J.~Phys.~Chem.~B {\bf 111}, 12017 (2007).
Copyright \copyright 2009 by the American Chemical Society.]
}
\label{fig:residue}
\end{figure}

We used the trajectory of cytochrome c in water 
generated by Bu and Straub \cite{BS03}.
To study the local nature of the amide I modes and 
the correspondence with experiment,
we isotopically labeled four specific CO bonds,
typically C$^{12}$O$^{16}$ as C$^{14}$O$^{18}$. 
In evaluating the potential energy in our instantaneous normal mode analysis, 
we truncated the system with an amide I mode at the center using
a cutoff ($\simeq 10$ \AA), including both protein 
and water.
Following INM analysis, 
we used Eq.~(\ref{eq:VER-1d}) to calculate the 
time course of the density matrix.
The predicted VER is single exponential in character with
time scales that are subpicosecond 
with relatively small variations induced 
by the different environments of the amide I modes
(see Fig.~\ref{fig:residue} and Table I in \cite{FS07} 
for numerical values of the VER timescales).
To identify the principal contributions to the dependence 
on the environment we examined
the VER pathways and the roles played by
protein and water 
degrees of freedom in VER. 
Our first conclusion is that, for the amide I modes 
buried in the protein ($\alpha$-helical regions),
the water contribution is less than for 
the amide I modes exposed to water (loop regions).
This finding is important because only a total 
VER timescale is accessible in experiment. 
With our method, the energy flow pathways into 
protein or water can be clarified.

\begin{figure}[b]
\begin{center}
\begin{minipage}{.42\linewidth}
\includegraphics[scale=1.2]{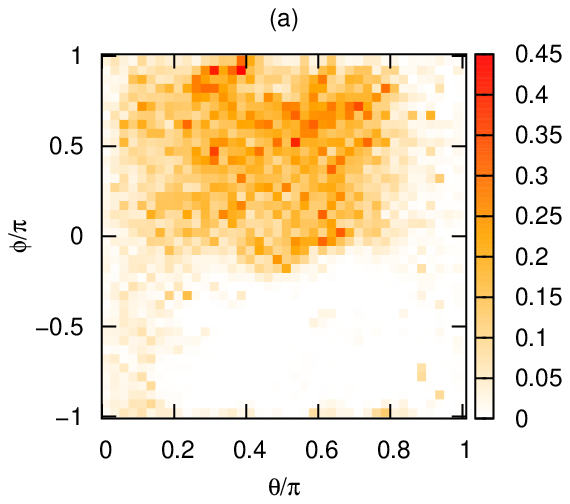}
\end{minipage}
\hspace{1cm}
\begin{minipage}{.42\linewidth}
\includegraphics[scale=1.2]{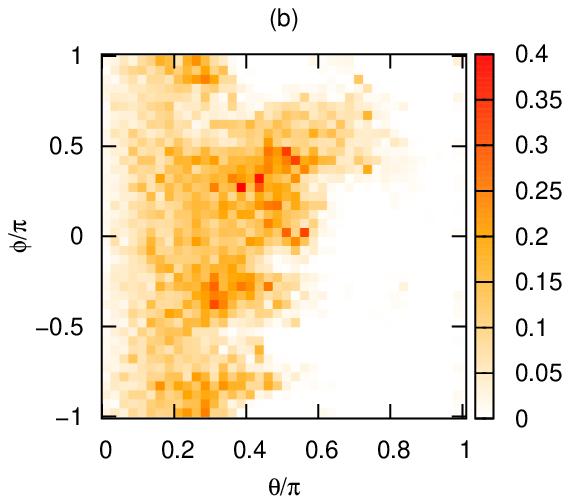}
\end{minipage}
\begin{minipage}{.42\linewidth}
\includegraphics[scale=1.2]{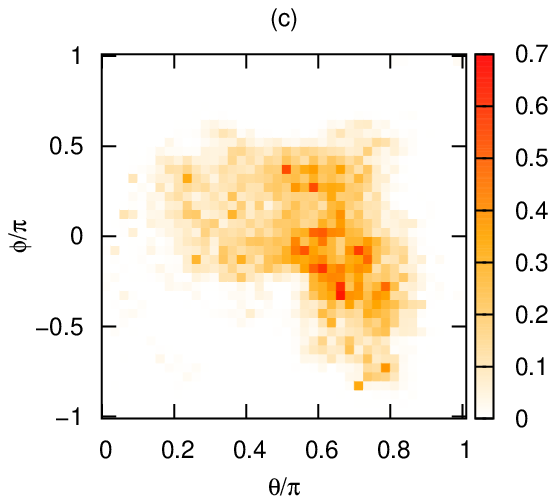}
\end{minipage}
\hspace{1cm}
\begin{minipage}{.42\linewidth}
\includegraphics[scale=1.2]{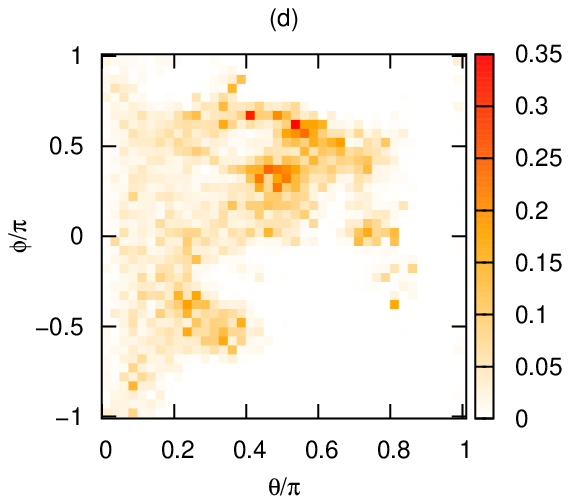}
\end{minipage}
\end{center}
\caption{\baselineskip5mm
Angular excitation functions for the resonant normal modes
of water
for the (a)  81st, (b) 84th, (c) 93rd, and (d) 97th residues, 
represented in arbitrary units.
[Reprinted with permission from 
H. Fujisaki and J.E. Straub,
J.~Phys.~Chem.~B {\bf 111},12017 (2007).
Copyright \copyright 2009 by the American Chemical Society.]
}
\label{fig:dist2-wat}
\end{figure}

Focusing on the resonant bath modes, we analyzed the anisotropy of the 
energy flow, as shown in Fig.~\ref{fig:dist2-wat},
where the relative positions of bath modes participating in VER 
are projected on the spherical polar coordinates ($\theta,\phi$) 
centered on the 
CO bond involved in the amide I mode, which represents the principal z-axis
(see Fig.~1 in \cite{FS07}). 
The angle dependence of the energy flow from the 
amide I mode to water is calculated from the normal mode 
amplitude average, and not directly related to experimental
observables.
As expected, energy flow is observed in
the direction of solvating water.
However, that distribution is not spatially isotropic 
and indicates preferential directed energy flow.
These calculations demonstrate the power of our theoretical analysis 
in elucidating pathways for spatially directed energy flow 
of fundamental importance to studies of energy flow 
and signaling in biomolecules 
and the optimal design of nanodevices (see summary and 
discussion for more detail).

\subsection{Porphyrin}
\label{sec:porphyrin}

Our final example is a modified porphyrin \cite{ZFS09a}.
We have carried out systematic studies of VER in the
porphyrin-imidazole complex, a system that mimics
the active site of the heme protein myoglobin (Mb). 
The structure of myoglobin
was first determined in 1958 \cite{Kendrew}. 
Experimental and computational studies 
exploring the dynamics of myoglobin 
led to the first detailed picture of how fluctuations in a protein structure 
between a multitude of ``conformational substates'' supports protein 
function \cite{Frauenfelder}.
Time resolved spectroscopic studies ~\cite{Hochstrasser}  
coupled with computational studies have provided a detailed picture 
of time scale and mechanism for energy flow in myoglobin 
and its relation to function. 
Karplus and coworkers developed the CHARMM force field \cite{CHARMM} 
for heme and for amino acids
for the study of myoglobin, 
and a particular focus on the dissociation and rebinding
of ligands such as CO, NO, and O$_2$ \cite{Mb}.
The empirical force field appears to provide an accurate model 
of heme structure and fluctuations. 
However, we have less confidence in the accuracy of anharmonicity 
essential to detailed mode coupling on the force field 
and the dependence on spin state, which is important to the proper identification 
of the electronic ground state potential energy surface.

\begin{figure}[b]
\hfill
\begin{center}
\includegraphics[scale=0.6]{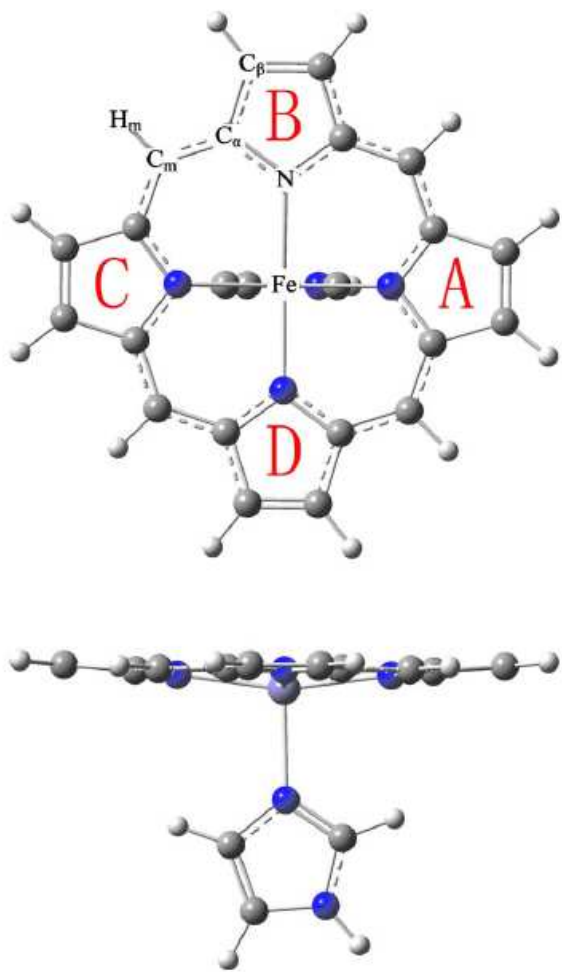}
\end{center}
\caption{\baselineskip5mm
Optimized structure of FeP-Im (quintuplet $S=2$ spin configuration) at 
the UB3LYP/6-31G(d) level of theory. 
[Reprinted with permission from 
Y. Zhang, H. Fujisaki, and J.E. Straub,
J.~Chem.~Phys.~{\bf 130}, 025102 (2009).
Copyright \copyright 2009 by the American Institute of Physics.]
}
\label{fig:porphyrin1}
\end{figure}

We carried out ab initio calculations 
for a heme-mimicking molecule, iron-porphine ligated to imidazole,
abbreviated as FeP-Im.
See Fig.~\ref{fig:porphyrin1} for the optimized structure.
We employed the UB3LYP/6-31G(d) level of theory as in the case of the 
isolated NMA \cite{FYHS07,FYSS09},
but carefully investigated the spin configurations.
We identified the quintuplet ($S=2$) as the electronic ground state,
in accord with experiment. 
Our study of VER dynamics 
on this quintuplet ground state potential energy surface (PES) is summarized here. 
Additional investigations of the VER dynamics on 
the PES corresponding to other spin configurations 
as well as different heme models
are described elsewhere \cite{ZFS09a}.

\begin{figure}[b]
\hfill
\begin{center}
\includegraphics[scale=0.6]{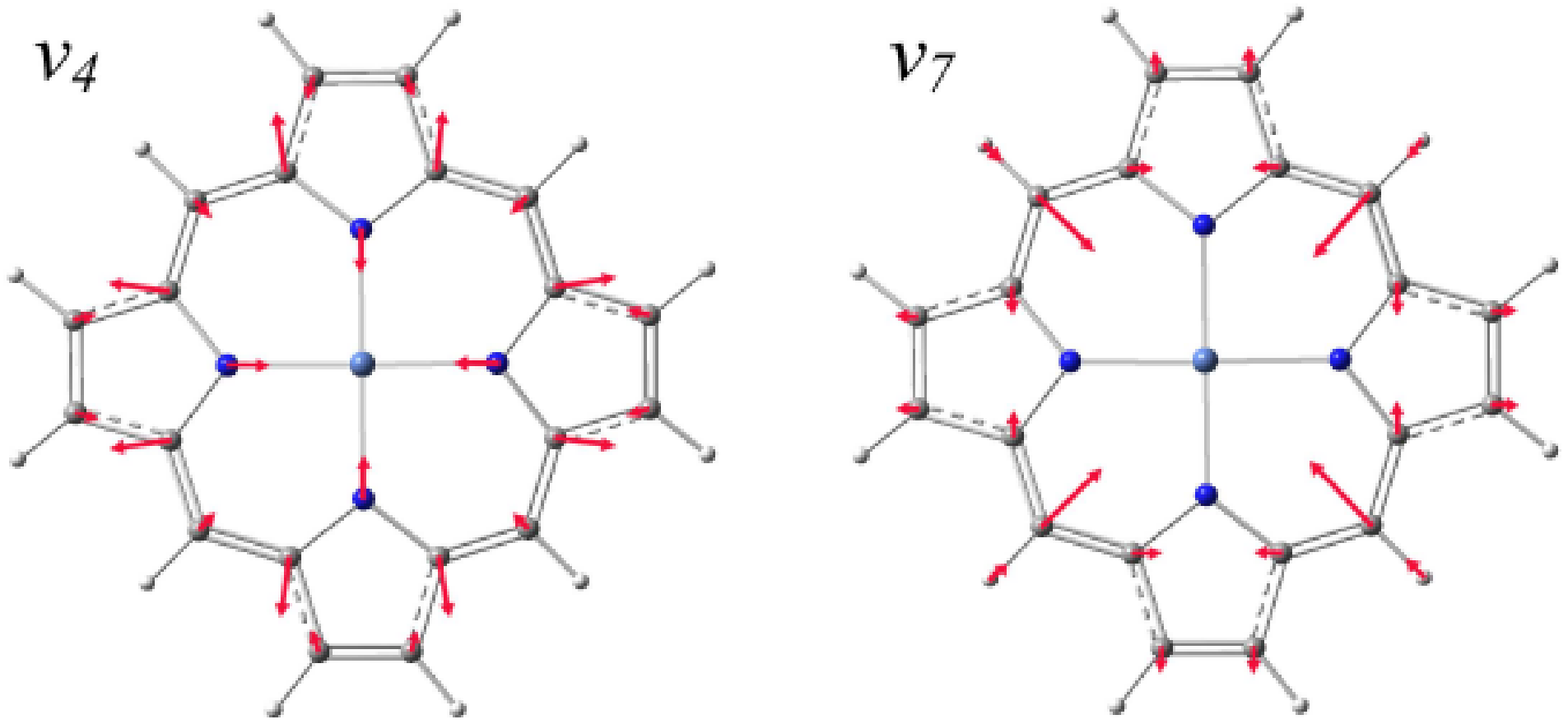}
\includegraphics[scale=0.6]{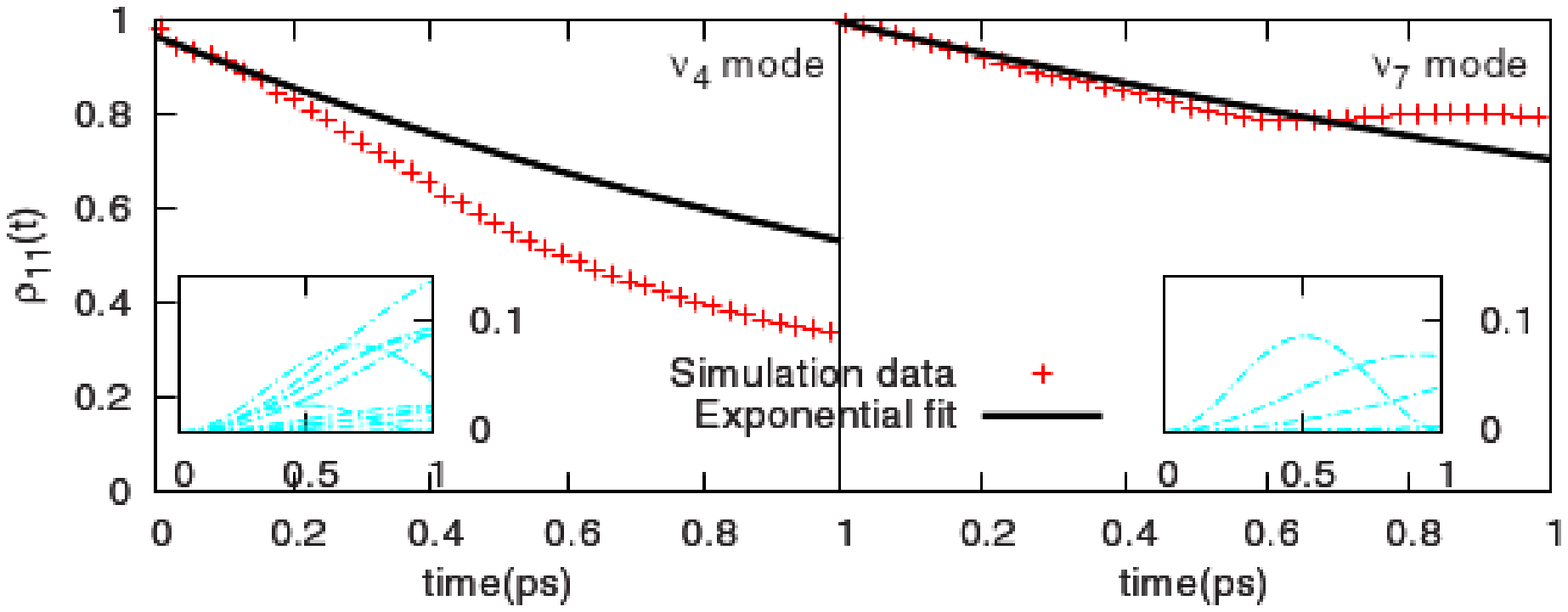}
\end{center}
\caption{\baselineskip5mm
Time course of the $\rho_{11}$ element of the reduced density matrix for 
$\nu_4$ and $\nu_7$ excitations of $v=1$.
[Reprinted with permission from 
Y. Zhang, H. Fujisaki, and J.E. Straub,
J.~Chem.~Phys.~{\bf 130}, 025102 (2009).
Copyright \copyright 2009 by the American Institute of Physics.]
}
\label{fig:porphyrin2}
\end{figure}

A series of elegant pioneering experimental studies
have provided a detailed picture of the dynamics of the
$\nu_4$ and $\nu_7$ modes, 
in-plane modes of the heme (see Fig.~\ref{fig:porphyrin2}), 
following ligand photodissociation in myoglobin.
Using time-resolved resonance Raman spectroscopy,
Mizutani and Kitagawa observed mode specific excitation and relaxation \cite{MK97,MK01}.
Interestingly, these modes
decay on different time scales.
The VER time scales are $\sim$ 1.0 ps for 
the $\nu_4$ mode and $\sim$ 2.0 ps for the $\nu_7$ mode.
Using a sub-10-fs pulse,
Miller and co-workers extended the range of the coherence 
spectroscopy up to 3000 cm$^{-1}$ \cite{Miller2}.
The heme $\nu_7$ mode was found to be most strongly excited
following $Q$ band excitation.
By comparing to the deoxyMb spectrum,
they demonstrated that the signal was derived from the structural transition 
from the six-coordinate to the five-coordinate heme.
Less prominent excitation of the $\nu_4$ mode was also observed.
The selective excitation of the $\nu_7$ mode,
following excitation of out-of-place heme doming,
led to the intriguing conjecture 
that there may be directed energy transfer of the heme excitation 
to low frequency motions connected to backbone displacement and to protein function.
The low frequency heme modes ($<$400 cm$^{-1}$) have been studied using
femtosecond coherence spectroscopy with a 50 fs pulse \cite{Champion}.
A series of modes at $\sim$40 cm$^{-1}$, $\sim$80 cm$^{-1}$, $\sim$130 cm$^{-1}$ and $\sim$170 cm$^{-1}$
were observed for several myoglobin derivatives.
The couplings between these modes were suggested.
It is a long term goal of our studies to understand, at the mode-specific level,
how the flow of excess energy due to 
ligand dissociation leads to the selective excitation of 
$\nu_4$ and $\nu_7$ modes. 

In our study, we ignore the transition in spin state 
that occurs upon ligand photodissociation 
and the associated electron-nuclear coupling 
that will no doubt be essential to an understanding 
of the ``initial state'' of the $\nu_4$ and $\nu_7$ 
vibrations following ligand photodissociation.
Our focus was on the less ambitious but important question of vibrational energy flow 
on the ground state ($S=2$) surface following excitation resulting from photodissociation.
We employed time-dependent perturbation theory, 
Eq.~(\ref{eq:VER-1d}), 
to model the mode-specific relaxation dynamics.
The initial decay process of each system mode was fitted
by a single-exponential function. The time constant of
$\simeq$ 1.7 ps was derived for the $\nu_4$ mode and $\simeq$ 2.9 ps for the
$\nu_7$ mode. These theoretical predictions, which make no assumptions
regarding the VER mechanism, agree well with previous
experimental results of Mizutani and Kitagawa for MbCO \cite{MK97}.

Vibrational energy transfer pathways were identified by
calculating the 3rd order Fermi resonance parameters, Eq.~(\ref{eq:Fermi}). 
For the excited $\nu_4$ and $\nu_7$ modes, the dominant 
VER pathways involve porphine out-of-plane motions as
energy accepting doorway modes. 
Importantly, no direct energy transfer
between the $\nu_4$ and $\nu_7$ modes was observed.
Cooling of the five Fe-oop (Fe-out-of-plane) modes, 
including the functionally important
heme doming motion and Fe-Im stretching motion,
takes place on the picosecond time scale. All modes dissipate
vibrational energy through couplings, weaker or stronger,
with low frequency out-of-plane modes involving significant
imidazole ligand motion. It has been suggested that these
couplings trigger the delocalized protein backbone motion,
important for protein function, which follows 
ligand dissociation in Mb. 

The $\gamma_7$ mode, a porphine methine wagging motion associated
with Fe-oop motion, is believed to be directly excited
following ligand photodissociation in MbCO. The coupling
of this mode to lower frequency bath modes is predicted to
be very weak. However, its overtone is strongly coupled to
the $\nu_7$ mode, forming an effective energy transfer pathway
for relaxation on the electronic ground state and excited state
surfaces. This strong coupling suggests a possible mechanism
of excitation of the $\nu_7$ mode through energy transfer
from the $\gamma_7$ mode. That mechanism is distinctly different
from direct excitation together with Fe-oop motion of the $\nu_4$
mode and supports earlier conjectures of mode-specific 
energy transfer following ligand dissociation in myoglobin
\cite{MK97,Miller2}.

\section{Summary and Discussion}
\label{sec:summary}

This chapter provides an overview of our recent 
work on the application 
of the non-Markovian theory of vibrational 
energy relaxation to a variety of systems of biomolecular 
interest, including protein backbone mimicking
amide I modes in $N$-methylacetamide
(in vacuum, in water cluster, and in solvent water)
amide I modes in solvated cytochrome c, 
and vibrational modes in 
a heme-mimicking porphyrin ligated to imidazole.
We calculated the VER time scales and mechanisms 
using Eq.~(\ref{eq:VER-time}), incorporating 
a fluctuating bath,
and Eq.~(\ref{eq:VER-1d}), using a static bath 
approximation,
and compared them to experiment when available.
The theory is based on the reduced model using 
normal mode concepts with 
3rd and 4th order anharmonicity, Eq.~(\ref{eq:reduced}). 
Applying the simple time-dependent perturbation theory,
and ensemble averaging the resulting density matrix, 
a non-Markovian theory of VER was obtained.
We extended the previous 
theory due to Fujisaki, Zhang and Straub \cite{FZS06} to more 
general situations where (1) the relaxing ``system'' has a 
multi-mode character and (2) when the system parameters 
depend on time \cite{FS08}. 
We also discussed the limitations of 
our VER formulas related to the assumptions 
upon which the earlier theories are based.

We are now in a position to discuss the future aspects 
of our work, and the connection to other biolomolecular 
systems or nanotechnological devices.

{\it \underline{Relation to enzymatic reaction}}: The role of vibrational 
motions in the mechanism of enzymatic reactions remains 
controversial \cite{Agarwal}.
In enzymology, the characterization of the enzymatic reaction 
rate is essential. Kinetic information is 
typically derived from substrate-enzyme kinetics experiments. 
In simulation, the free energy 
calculation combined with transition state theory is 
the most powerful and practical way to compute reaction
rates. As enzymatically catalyzed reactions 
typically involve chemical bond 
breaking and formation, QM/MM type methods 
should be employed. Warshel and coworkers have examined this issue for 
several decades, and concluded that characterizing the 
free energy barrier is the most important consideration, 
noting that the electrostatic 
influence from the protein (enzyme) plays a key role \cite{Warshel}.
However, Hammes-Schiffer and coworkers have identified important situations 
in which VER might play a role in controlling the rate of 
enzymatic reactions \cite{Agarwal}.
Furthermore, Hynes and coworkers applied the Grote-Hynes theory to the 
enzymatic reactions, and investigated the dynamic role of 
the environment \cite{PTMHR08}. 
These recent studies indicate the importance of incorporating 
vibrational energy flow and dynamics as part of 
a complete understanding of enzyme kinetics.

{\it \underline{Relation to conformational change}}:
The relation between vibrational excitation/relaxation and 
conformational change of molecules is intriguing in part 
because of the possible relation to the optimal control of 
molecular conformational change using tailored laser pulses. 
It is well known that there are dynamic corrections 
to the RRKM reaction rate: the simplest being
\begin{equation}
k(E)=k_{RRKM}(E)/(1+\nu_R/k_{IVR}(E))
\end{equation}
where $\nu_R$ is the intrinsic frequency of a reaction 
coordinate,
$k_{IVR}(E)$ is a microcanonical IVR rate, 
and 
$k_{RRKM}(E)$ is the RRKM reaction rate \cite{Steinfeldbook,Billingbook,BBS88}.
Several modifications to this formula
are summarized in \cite{Leitner}.
It is obvious that VER affects how a molecule changes its shape.
However, this is a ``passive'' role of VER. 
Combining RRKM theory and the local random matrix theory \cite{LW97},
Leitner, Wales and coworkers theoretically studied the active role 
of vibrational excitations on conformational change
of a peptide-like molecule (called NATMA) \cite{ALEW05}.
There are two particular modes (NH stretching) in NATMA, 
and they found that the final product depends on 
which vibrational mode is excited \cite{DLWZ04}. 
For the same system, Teramoto and Komatsuzaki further refined 
the calculation by employing ab initio potential energy surface \cite{TK09}.
A possibility to control molecular configurations of peptides or proteins
using laser pulses should be pursued and 
some experimental attempts have begun \cite{Hamm08,Miller,Zwier}.

Another interesting attempt should be to address  
mode specific energy flow associated with structural change.
Recently Ikeguchi, Kidera, and coworkers \cite{IUSK05}
developed a linear response theory for conformational changes of 
biomolecules, which is summarized in another chapter 
of this volume \cite{FMK09}.
Though the original formulation is based on a static 
picture of the linear response theory (susceptibility),
its nonequilibrium extension may be used to explore 
the relation between energy flow and conformational change 
in proteins.
In addition,
Koyama and coworkers \cite{Koyama08} 
devised a method based on principal component analysis 
for individual interaction energies of a peptide (and water), 
and found an interesting correlation between the principal modes 
and the direction of conformational change \cite{Koyama08}.

{\it \underline{Relation to signal transduction in proteins}}:
Though signal transduction in biology mainly denotes 
the information transfer processes carried out 
by a series of proteins in a cell, 
it can be interesting and useful to study the information 
flow in a single protein, which should be related to 
vibrational population and phase dynamics.
Straub and coworkers \cite{Straub} studied such energy flow pathways 
in myoglibin, and found particular pathways from heme to 
water, later confirmed experimentally by
Kitagawa, Mizutani and coworkers \cite{KNM06} 
and Champion and coworkers \cite{Champion}.
Ota and Agard \cite{OA05} devised a novel simulation protocol, 
{\it anisotropic thermal diffusion}, 
and found a partitular energy flow pathway 
in the PDZ domain protein. Importantly, 
the pathway they identified is located near the 
conserved amino acid region in the protein family previously
elucidated using information theoretic approach 
by Lockless and Ranganathan \cite{LR99}.
Sharp and Skinner \cite{SS06} proposed an alternative method, 
{\it pump-probe MD}, and examined the 
same PDZ domain protein, identifying alternative
energy flow pathways. Using linear response theory describing
 thermal 
diffusion, Ishikura and Yamato \cite{Yamato} 
discussed the energy flow 
pathways in photoactive yellow protein. 
This method was recently extended 
to the frequency domain by Leitner and  
applied to myoglobin dimer \cite{Leitner09b}.
Though the energy flow mentioned above occurs quite rapidly ($\sim$ ps), 
there are time-resolved spectroscopic methods to 
detect these pathways in vitro \cite{Hamm08}.
Comparison between theory and experiment 
will help clarify the biological role
of such energy flow in biomolecular systems. 

{\it \underline{Exploring the role of VER in nanodevice design}}:
Applications of the methods described in this chapter 
are not limited to 
biomolecular systems. 
As mentioned in the introduction, heat generation is 
always an issue in nanotechnology, 
and an understanding of VER in molecular devices 
can potentially play an important role in optimal device design.
The estimation of thermal conductivity in such devices 
is a good starting point recently pursued 
by Leitner \cite{Leitner}. 
Nitzan and coworkers studied thermal conduction in a molecular wire 
using a simplified model \cite{SNH03}. It will be interesting
to add more molecular detail to such model calculations.
Electronic conduction has been one of the main topics in
nanotechnology and mesoscopic physics \cite{Cuniberti}, and heat 
generation during electronic current flow is an additional 
related area of importance.

{\it \underline{VER in a confined environment}}:
We have found evidence for spatially anisotropic 
vibrational energy flow with specific pathways 
determined by resonance and coupling conditions.
It was shown for amide I modes in cytochrome c 
that VER may depend on the position 
of the probing modes \cite{FS07}, 
making it useful for the study of
inhomogeneity of the environment.
For example, an experimental study of VER 
in a reverse micelle environment \cite{ZBO03}, 
fullerene, nanotube, membrane, or 
on atomic or molecular surfaces \cite{Saalfrank}
may all be approached using methods described in this chapter.

{\it \underline{Anharmonic effects in coarse-grained models 
of proteins}}:
Recently Togashi and Mikhailov studied the conformational relaxation 
of elastic network models \cite{Togashi}. 
Though the model does not explicitly incorporate
anharmonicity, small anharmonicity exists, resulting 
in interesting physical behavior relevant to biological function. 
Sanejouand and coworkers 
added explicit anharmonicity into the elastic network
models, and studied the energy storage \cite{Sanejouand} 
through the lens of ``discrete breather'' ideas from nonlinear science \cite{discrete}. 
Surprisingly, they found that the energy storage may occur in the active 
sites of proteins. It remains to be seen whether their conjecture
will hold for all-atom models of the same system.


\begin{acknowledgments}
The authors gratefully acknowledge fruitful and enjoyable collaborations 
with Prof.~G. Stock, Prof.~K. Hirao, and Dr.~K. Yagi.
The results of which form essential contributions to this chapter.
We thank Prof. David M. Leitner, Prof. Akinori Kidera, Prof. Mikito Toda, 
Dr. Motoyuki Shiga, Dr. Sotaro Fuchigami, Dr. Hiroshi Teramoto for useful discussions.
The authors are grateful for the generous support of this research by 
the National Science Foundation (CHE-0316551 and CHE-0750309) 
and Boston University's Center for Computer Science.
This research was supported by Research and Development of 
the Next-Generation Integrated Simulation of Living Matter,
a part of the Development and Use of the Next-Generation 
Supercomputer Project of the Ministry of Education,
Culture, Sports, Science and Technology (MEXT).

\end{acknowledgments}







\end{document}